\documentclass[a4paper,11pt]{article}
\pdfoutput=1 

\usepackage{jcappub} 
\usepackage[T1]{fontenc} 
\usepackage{array}
\usepackage{booktabs}
\usepackage{enumitem}
\usepackage{framed}
\usepackage{mathtools}
\usepackage{lipsum}
\usepackage{xcolor}

\newcommand{\omatter}{\ensuremath{\Omega_{\mathrm{m}}}}
\newcommand{\wzero}{\ensuremath{w_{0}}}
\newcommand{\hubble}{\ensuremath{H_{0}}}
\newcommand{\ns}{\ensuremath{n_{s}}}
\newcommand{\obary}{\ensuremath{\Omega_{b}}}
\newcommand{\sigeight}{\ensuremath{\sigma_8}}

\newcommand{\Seight}{\ensuremath{S_8}}
\newcommand{\wcdm}{\ensuremath{w}CDM}
\newcommand{\x}{\ensuremath{\times}}
\newcommand{\vir}{\ensuremath{_{\mathrm{200}}}}
\newcommand{\cosmogrid}{\mbox{\textsc{CosmoGridV1}}}
\newcommand{\pycosmo}{\mbox{\textsc{PyCosmo}}}
\newcommand{\pkdgrav}{\textsc{PkdGrav3}}
\newcommand{\ufalcon}{\textsc{UFalcon}}
\newcommand{\concept}{\textsc{CO}\emph{N}\textsc{CEPT}}
\newcommand{\nside}{{\texttt{nside}}}
\newcommand{\lbox}{\ensuremath{L_{\rm{box}}}}
\newcommand{\npart}{\ensuremath{n_{\rm{part}}}}
\newcommand{\cl}{\ensuremath{C_{\ell}}}
\newcommand{\npeaks}{\ensuremath{n_{\mathrm{peaks}}}}

\newcommand{\rev}[1]{#1}
\defcitealias{Fluri2019kids}{F19}
\defcitealias{Fluri2022kids}{F22}
\defcitealias{Zuercher2022despeaks}{Z22}

\notoc

\title{\boldmath CosmoGridV1: a simulated $w\rm{CDM}$ theory prediction for map-level cosmological inference}

\author[1]{Tomasz Kacprzak,}
\author[1]{Janis Fluri,}
\author[2]{Aurel Schneider,}
\author[1]{Alexandre Refregier,}
\author[2]{Joachim Stadel}

\affiliation[1]{ETH Zurich, Wolfgang-Pauli-Str 27, 8093 Zurich, Switzerland}
\affiliation[2]{University of Zurich, Winterthurerstrasse 190, 8057 Zurich, Switzerland}

\emailAdd{tomaszk@phys.ethz.ch}
\emailAdd{jafluri@phys.ethz.ch}

\abstract{
We present \cosmogrid: a large set of lightcone simulations for map-level cosmological inference with probes of large scale structure.
It is designed for cosmological parameter measurement based on \mbox{Stage-III} photometric surveys with non-Gaussian statistics and machine learning.
\cosmogrid\ spans the \wcdm\ model by varying \omatter, \sigeight, \wzero, $H_0$, \ns, \obary, and assumes three degenerate neutrinos with fixed \mbox{$\sum m_\nu$ = 0.06 eV}. 
This space is covered by 2500 grid points on a Sobol sequence.
At each grid point, we run 7 simulations with \textsc{PkdGrav3} and store 69 particle maps at \nside=2048 up to $z$=3.5, as well as halo catalog snapshots.
The fiducial cosmology has 200 independent simulations, along with their stencil derivatives. 
An important part of \cosmogrid\ is the benchmark set of 28 simulations, which include larger boxes, higher particle counts, and higher redshift resolution of shells.
They allow for testing if new types of analyses are sensitive to choices made in \cosmogrid. 
We add baryon feedback effects on the map level, using shell-based baryon correction model.
The shells are used to create maps of weak gravitational lensing, intrinsic alignment, and galaxy clustering, using the \textsc{UFalcon} code.
The main part of \cosmogrid\ are the raw particle count shells that can be used to create full-sky maps for a given $n(z)$.
We also release projected maps for a Stage-III forecast, as well as maps used previously in KiDS-1000 deep learning constraints with \cosmogrid.
The data is available at \href{https://www.cosmogrid.ai}{\url{www.cosmogrid.ai}}.
}

\begin{document}
\maketitle
\flushbottom

\section{Introduction} \label{sec:intro}

The patterns and structures present in the matter density field carry cosmological information about the composition and history of the universe, as well as the laws of physics governing its evolution.
Multiple large scale structure (LSS) probes, such as weak gravitational lensing or galaxy clustering, are used to make measurements of cosmological parameters within specified cosmological models \citep{Des2022combined,Heymans2020multiprobe,Hikage2019cosmicshear}.
The LSS is the most effective at constraining the matter density \omatter\ and matter clustering amplitude \sigeight.
Moreover, as the LSS maps can be created in tomographic bins, we can trace the evolution of the matter density field over cosmic time; this enables measurements of the dark energy equation of state \wzero\ and its evolution, which can bring us closer to understanding the nature of the cosmic acceleration.

Recent measurements from the Dark Energy Survey\footnote{\url{www.darkenergysurvey.org}} (DES), the Kilo-Degree Survey\footnote{\url{kids.strw.leidenuniv.nl}} (KiDS), and the Hyper-Suprime Cam\footnote{\url{hsc.mtk.nao.ac.jp/ssp}} (HSC) have measured these parameters with $<5\%$ precision. 
The measurements of the $S_8=\sigma_8(\Omega_m/0.3)^{0.5}$ parameter from these surveys indicate a mild tension with the value obtained from the Cosmic Microwave Background \citep[CMB,][]{Planck2018parameters,Amon2022consistent,Leauthaud2017lensinglow,Leauthaud2022withoutborders,cosmology2022intertwined}.
Upcoming experiments, such as the Legacy Survey of Space and Time of the Rubin Observatory\footnote{\url{www.lsst.org}} and Euclid\footnote{\url{www.euclid-ec.org}} are going to improve these measurements, breaking the 1\% precision threshold  \citep{Albrecht2006darkenergy,Amendola2018Euclid,Zhan2018lsst}.
However, before the data from these Stage-IV experiments becomes available, the existing datasets will be studied further in different ways to shed more light on the \Seight\ tension.
A method that recently has gained interest is the simulations-based inference, where the theory prediction for the observed maps is created directly from simulations.
Map-level analysis can extract more information from the same dataset than the classical two-point function analysis, respond differently to systematic errors, as well as break some key degeneracies between the model parameters.

\subsection{Inference with non-Gaussian features} \label{sec:non_gaussian_inference}

On large scales, the density fluctuations are well represented by a Gaussian Random Field, for which the 2-pt functions are a sufficient statistic.
On intermediate and small scales, the effects of gravitational interactions throughout the cosmic time give raise the non-Gaussian structures in the matter density field; forming a complex network of halos, filaments, sheets and voids, known as the \emph{cosmic web} \citep{bond1996filaments,Coles2000characterizing,Dietrich2012filament}.
In a conventional analysis, the two-point functions are used as the summary statistic of choice to compare the survey measurements with theory prediction.
These statistics, such as the angular power spectra, real space correlation functions, or wavelet-like COSEBIs \citep{asgari2020kids,Kilbinger2015review} capture the Gaussian information contained in the maps.
While they extract all the available information at large scales, they miss the non-Gaussian information content at intermediate and small scales.
The smaller the scales used, the more information is contained in the  non-Gaussian features \citep{Fluri2018deep,Gupta2018,zuercher2020forecast}.

Various features have been proposed to extract the information beyond 2-pt, including the bispectrum \citep{sefusatti2006cosmology,dodelson2005weak,fu2014cfhtlens,rizzato2019tomographic} and trispectrum \citep{munshi2021weak},
higher order moments of mass maps \citep{takada2002kurtosis,vafaei2009breaking,Patton2017convergence1pt,gatti2021desyear3}, 
Minkowski functionals \citep{pratten2012nongaussianity,kratochvil2011probing,matilla2017geometry,Liu2022minkowski},
weak lensing voids \citep{Davies2021void,Davies2019voids}, and wavelet decomposition coefficients \cite{Ajani2021starlet}.
The feature that has been most extensively studied is the shear peaks counts \citep{DietrichHartlap2010peaks,Liu2016a,Fluri2018peaks,zuercher2020forecast,Ajani2020peaks,zurcher2022towards}
The peak counts were used to make measurements from surveys \citep{Liu2014,liu2014cosmology,Kacprzak2016peaks,Martinet2017,harnoisDeraps2021desy1} and recently by \citet[][hereafter \citetalias{Zuercher2022despeaks}]{Zuercher2022despeaks}.
Several non-Gaussian analyses combine features from different probes: the density split statistics \citep{Gruen2018densitysplit,Friedrich2018density,Burger2022kids1000}, and Minkowski functionals \citep{grewal2022minkowski}.
Finally, the non-Gaussian statistics are often combined with the 2-pt to further increase the constraining power \citep{Berge2010optimal,Zuercher2022despeaks,liu2014cosmology}.

Recently, machine learning methods have been proposed to automatically design features that maximize the information extracted from a given dataset.
For the LSS, the most studied method is deep learning, using convolutional neural networks \citep[CNN,][]{Schmelzle2017,Fluri2018deep,Gupta2018,Ribli2019cnn,Fluri2021inference,KacprzakFluri2022deeplss}.
In this paradigm, the CNNs are typically trained to create features that maximize the discriminating power of the network between the cosmological parameters. 
Most often, these features are then interpreted in a likelihood analysis, by creating a conditional likelihood function of the features given the truth input from simulations \citep{Gupta2018,Fluri2018deep}; this way the features are treated the same way as for the conventional 2-pt analysis, where the \cl\ or $\xi_{\pm}$ is the pre-defined feature vector.
Recently, \citet{KacprzakFluri2022deeplss} proposed a deep learning combined probes analysis that includes lensing and clustering simultaneously.
Measurements using deep learning from lensing maps have been performed by \citet{Fluri2019kids,Fluri2022kids}, where the CNN method obtained greater constraining power than the power spectra alone.

Regardless of the summary statistic used, the benefits of performing a map-level, simulation-based analysis of the LSS data can be as follows:
\begin{itemize}
    \item extracting non-Gaussian information and thus improving the precision of measurements from the same dataset, or alternatively decreasing the dependence on small scales to obtain a measurement with comparable precision \citepalias{Zuercher2022despeaks},
    \item beyond 2-pt analysis can break key degeneracies between the parameters of assumed model and/or systematics \citep{Peel2018degeneracies,KacprzakFluri2022deeplss,Lu2022baryonic},
    \item machine learning analysis can efficiently localize the features in maps of different probes and increase the signal-to-noise by ignoring the uninformative regions \citep{KacprzakFluri2022deeplss},
    \item full forward modelling of the observed maps can naturally include various observational effects, such as noise, survey mask, systematics, and others, which can simplify the analysis (see section below).
\end{itemize}

\rev{Simulations-based inference comes with a number of unique challenges.
Firstly, there still exist small differences between N-body simulation engines, which produce slightly different maps if started from the same initial conditions \citep{Schneider2016challenge,AnguloHahn2022review}.
The impact of these differences on map-level inference will be studied in future work.
Secondly, the uncertainty over the baryonic feedback has been mostly studied for 2-pt functions \cite{Mead2021HMcode,Schneider2020baryons1}, but more work is needed to understand its impact on map-level inference.
Finally, the survey systematics testing framework has been also been mostly designed for 2-pt functions. 
While methodologies for testing the impact of PSF calibrations and redshift errors have been demonstrated for the peak counts \cite{Kacprzak2016peaks,Zuercher2022despeaks}, more work is needed to develop it for other statistics.}

\subsection{Map-level theory prediction using simulations} \label{sec:theory_prediction}

The non-linear evolution of the density field is simulated using the N-body technique \citep[see][for review]{AnguloHahn2022review}.
In the classical 2-pt analysis, large N-body simulations with realistic galaxy population properties \citep{potter2016pkdgrav3,DeRose2019buzzard,Fosalba2015mice,nelson2015illustris} have become indispensable in several key areas, such as (among others):
(i) predicting the non-linear part of the matter density power spectrum $P(k)$, whether through the halo model formalism \citep{Takahashi2012halofit}, or directly \citep{knabenhans2019euclid,Lawrence2017cosmicemu}, 
(ii) for validation and testing of photometric redshift measurements \citep{myles2021dark,hildebrandt2021kids},
(iii) calculating or testing covariance matrices \citep{HarnoisDeraps2018simulations,shirasaki2019mock},
(iv) validating the end-to-end inference pipelines \citep[][and others]{MacCrann2018validating}.

In the map-level analysis using non-Gaussian features or machine learning, the simulations are used directly as the theory prediction.
Typically, a large suite of N-body simulations is created and processed to give the desired probe maps.
The observational effects, such as noise, survey masks or measurement systematics, are naturally added to the simulations in the forward modelling process.
These simulations can be created on a fixed set of grid points \citep{DietrichHartlap2010peaks,zuercher2020forecast}, or during the inference process itself, in an \emph{active learning} framework \citep{Alsing2019lfi}.
The grid simulations typically store the full dataset and post-process it later to create the forward-modeled maps of a target survey.
The simulations can be ran in \emph{snapshot} or \emph{lightcone} mode.
The former requires to store the full particle positions at a set of given timesteps,  the latter is based on thin particle shells at the given cosmic redshift \citep{Fosalba2008onion,Fosalba2015mice}.
After the simulations are ran, a set of tomographic survey maps is created using either the Born approximation or direct ray tracing, \citep[see][for more details]{Petri2017born,Tian2022lightcone}.
These maps are created by integrating the shell particle density against the relevant probe kernel, which is typically a function of the redshift distribution of selected galaxy sample $n(z)$ and cosmological parameters.

The key challenge in the simulation-based inference approach is its large computational effort.
The N-body simulations are computationally expensive, as they require computing the evolution of billions of particles in very small time steps.
Several codes have been developed and highly optimized to make these simulations fast \citep{HarnoisDeraps2013cube3pm,potter2016pkdgrav3,Habib2015hacc,Ragagnin2020gadget3,Springel2021Gadget4}.
Moreover, \emph{box replication} schemes are often employed to create a larger simulation with higher particle density, but with the same repeating particle distribution, taking advantage of the periodic boundary conditions.
Depending on a configuration, the runtime of a single simulation can take between the order of days to months.
Aside from runtime, data storage poses another challenge. 
A single snapshot or lightcone can take $\mathcal{O}(100)$ gigabytes.
The effects of these choices have been studied by \citet{Matilla2020optimizing,Sgier2021combined}.
Finally, the choice of physics included in the simulations will have a dominating influence over the simulations runtime. 
Different hydrodynamical models can be used for simulating small scale baryonic effects with varying level of precision.
In the order of increasing realism and computational cost, they can be categorized into (i) halo model-based \emph{baryonification} schemes \citep{Schneider2015baryons,Schneider2020xray,Arico2020baryons}, (ii) subgrid models of stellar and Active Galactic Nucleus (AGN) \citep{Brun2013yva,McCarthy2017bahamas}, or (iii) high resolution physical models relevant for galaxy formation, including AGN and black holes \citep{Sijacki2015Illustris,VillaescusaNavarro2020camels}. 

In creating a simulation set designed for a map-level inference with a target dataset, the following factors need to be considered.

\paragraph{The number of grid points within the prior space.} A small number of grid point will not sample the large volume of high-dimensional parameter space well enough, which will lead to interpolation errors of the signal between the sampled points. 

\paragraph{Simulation box size.} If too small, multiple replicated boxes will be needed, leading to underestimation of variance in the signal, similarly as in super-sample covariance terms in the 2-pt analysis \citep{Li2014supersample,Beauchamps2022supersample}. Furthermore, the results can be biased due to the missing of large modes \citep{Mohammed2014analytic,Schneider2016challenge}.

\paragraph{Simulation particle count.} If too small, the created maps will model the small scales incorrectly and contain larger shot noise, which may dominate the measurement errors of the survey.

\paragraph{Thickness of lightcone shells.} The maps created by integrating the lightcone by with the $n(z)$-dependent kernels may not be precise enough if the thickness of the shell is too large compared to the $n(z)$ of the tomographic bin.

\paragraph{\rev{Number of independent simulations.}} A large number of simulations from different initial conditions is needed to capture the cosmic variance. The error on the cosmic variance contribution to the likelihood function, whether created using a single covariance matrix or with conditional density estimation, needs to be subdominant compared to the survey statistical and systematic errors. If the total number of simulations is too small, the errors in cosmic variance will dominate the measurement error budget \citep{Petri2016samplevariance,Sellentin2016estimated}.

\paragraph{Model for baryonic physics} this choice depends on targeted length scales, which in turn are limited by the systematic errors of the analyzed dataset. The baryon physics model must be realistic enough model the chosen scales accurately. This choice is also coupled with particle count.

\vspace{1em}
Given the computational power and storage limitations, trade-offs often have to be made in choosing the simulations parameters.  
These choices are typical made for a specific scientific goal of the project, which include the cosmological parameters to be constrained, the noise levels of the data for the probes considered, the angular size of the survey, and the impact of systematics.

\subsection{Available simulation suites for map-level inference} \label{sec:available_sims}

There has been a number of simulation campaigns aimed at different types of analyses with 2-pt functions or non-Gaussian features.
In this section we will shortly review the simulation grids that already have been used for map-level parameter inference from large scale structure probes, either in a measurement from survey data or in a forecast.
We consider only the simulations that have the lightcone shells/snapshots output dense enough in the redshift space to enable map-level inference with projected probe maps from LSS surveys.

The first simulations set of this kind was created by \citet{DietrichHartlap2010peaks}, and was used for the peak counts cosmology with Dark Energy Survey \citep{Kacprzak2016peaks} and KiDS-450 \citep{Martinet2017}.
\mbox{\textsc{MassiveNus}} \citep{Liu2018massiveNus}, a dataset aimed at constraining neutrino masses with LSS data, used a pencil-beam approach to building a lightcone to generate $\mathcal{O}$(10000) realizations of 3.5$\times$3.5 deg$^2$ convergence maps by randomly rotating and shifting the simulation boxes.
It was used in multiple forecasts \citep{Liu2022minkowski,Ajani2020peaks,cheng2021scattering}.
The state-of-art simulations \textsc{cosmo-SLICS} \citep{HarnoisDeraps2019cosmoslics} were used in a many papers, notably for the DES-Y1 analysis beyond the 2-pt, with shear peak statistics.
In that work, flat patches of 10$\times$10 deg$^2$ were used.
The \textsc{DarkGridV1} simulation set is the small precursor to \cosmogrid, and was used in a DES-Y3 combined shear peak statistics and power spectrum analysis by \citetalias{Zuercher2022despeaks}.
It created maps on the sphere using the box replication approach, at the Healpix \nside=1024.
\cosmogrid\ was used in the KIDS-1000 deep learning analysis by \citet{Fluri2022kids} and is currently being used for further DES projects.
Table~\ref{tab:sims_compare} shows the comparison of these simulations in terms of number of grid points, box sizes, and particle counts.
We defer the detailed comparison of \cosmogrid\ and other simulations sets to the conclusions section.

The presented list does not include simulations that are designed for other science objectives and may not be immediately suitable for map-level inference. 
Large simulations, such as \textsc{Quijote} \citep{VillaescuseNavarro2020quijote}, \textsc{Camels} \citep{VillaescusaNavarro2020camels}, \textsc{Abacus} \citep{Maksimova2021abacussummit}, \textsc{Indra} \citep{Falck2021indra}, \textsc{Bacco} \citep{Angulo2021bacco}.
and others, also produce simulations spanning multiple cosmological parameters, but store the snapshots at a small number of redshifts, and do not anticipate the production of projected probe maps for any suvery's $n(z)$.
They can still be used for inference using pre-defined non-Gaussian features, such as halo mass functions, but not straightforwardly in a forward-modelling, map-based framework.

\subsection{The public \cosmogrid\ simulation set} \label{sec:cosmogrid_intro}

In this paper we present \cosmogrid: a large lightcone simulation set for map-level, simulation-based cosmological inference with probes of large scale structure.
\cosmogrid\ is designed for practical parameter measurement with the Stage-III survey data, for example with KiDS, DES, and HSC.
It contains 2500 unique cosmological parameters spanning the \wcdm\ model, with 7 unique simulations at each point. 
Additionally, there are 200 unique simulations at the fiducial cosmology, each with two derivative steps $\pm \Delta$ for each of the 6 parameters.
The total number of independent N-body simulations is 17500+200.
Including the fiducial derivatives, that leads to the total of 20100 simulations.
It also contains a set of benchmark simulations, which allow for validating the analysis choices, such as the feature vector (peaks, Minkowski functionals, machine learning features, and others), against the limitations of the chosen main configuration of the simulations, sich as the box size, number of particles, and shell thickness.

\cosmogrid\ also enables fast shell-level addition of baryonic feedback effects, using the baryonification scheme by \citet{Schneider2019baryon}.
To achieve this, we store friends-of-friends halo catalogs for each time step, with minimum particle count of 150.
Intrinsic alignments and galaxy biasing models can be added during post-processing.

\colorlet{shadecolor}{blue!7.5}
\begin{shaded}
\noindent
The \cosmogrid\ dataset consists of a total of 20128 simulations divided into three main parts: 
\begin{itemize}
\setlength\itemsep{0em}
\item {\bf grid}:   a set of 2500 cosmologies, each with 7 simulations from unique initial conditions (a total of 17500 N-body runs), 
\item {\bf fiducial}: simulations and the fiducial cosmology and its $\pm\Delta$ derivatives, with 200 unique initial conditions (2600 runs),
\item {\bf benchmark}: simulation benchmarks used for systematics testing of features chosen for parameter inference (28 runs).
\end{itemize}
The data is hosted at \href{https://www.cosmogrid.ai}{\url{www.cosmogrid.ai}} by ETH Zurich and is available via the Globus transfer. For each simulation, we store:
\begin{enumerate}
\setlength\itemsep{0em}
\item raw simulation lightcone particle count maps stored at Healpix \nside=2048 up to \mbox{$z<3.5$} (up to 69 shells per simulation),
\item halo catalog snapshots created using the friends-of-friends halo finder, with halo mass of \mbox{$M \sim 10^{13} \ h^{-1} M_{\odot}$}, together \rev{with fitted NFW parameters}, at every time step,
\item projected full sky weak lensing, galaxy density, and intrinsic alignments maps for a Stage-III forecast, including baryonification, at the \nside=512,
\item projected KiDS-1000 lensing and intrinsic alignment maps with grid extended with baryonic feedback parameters, from \citet{Fluri2022kids}.
\end{enumerate}
\end{shaded}

This paper accompanies the public release of the \cosmogrid\ simulations suite.
We present the simulations sets and motivate the choices made in designing them.
Furthermore, we demonstrate the methodology that can be employed for testing the analysis choices using the benchmark simulations.
We do this, we show an an example mock analysis of convergence and galaxy density maps that uses power spectra and peak statistics.

The simulations were created at the Swiss Supercomputing Center\footnote{\url{www.cscs.ch}} (CSCS) within the large production proposal called ``Measuring Dark Energy with Deep Learning'', which ran in 2020 - 2021.
The simulation code used was \pkdgrav\ \citep{potter2016pkdgrav3}, which ran on the Piz Daint cluster with GPU acceleration. 
The total computing time used to create this simulation set was approximately 750'000 GPU node hours.

This paper is organized as follows.
In Section~\ref{sec:grid} we describe the choice of the simulated cosmologies for the grid and fiducial parameters.
Section~\ref{sec:sims} describes the dark matter simulations and motivates the choices of N-body parameters.
We describe the procedure used to create survey maps in Section~\ref{sec:maps}, including the novel \emph{shell permutation} scheme. 
The inclusion of baryon feedback is described in Section~\ref{sec:baryons} and intrinsic galaxy alignments in Section~\ref{sec:intrinsic_alignments}.
We show the power-spectrum level validation tests for the projected maps in Section~\ref{sec:tests}.
We demonstrate how to use the benchmark simulations to validate the feature vector choices in Section~\ref{sec:benchmarks}.
We conclude in Section~\ref{sec:conclusions}.

\begin{table*}
\begin{center}
\setlength{\tabcolsep}{4pt}
\footnotesize{\def\arraystretch{1.2}
\begin{tabular}{
 >{\raggedright\arraybackslash} p{2.6cm} 
 >{\raggedright\arraybackslash} p{4.0cm}  
 >{\raggedright\arraybackslash} p{1.8cm}  
 >{\raggedright\arraybackslash} p{2.2cm}  
 >{\raggedright\arraybackslash} p{1.3cm}  
 >{\raggedright\arraybackslash} p{1.3cm}  
 }
\toprule
                        & Variable parameters                                                   & Number of simulations for variable parameters      & Number of simulations for the fiducial cosmology   & Box size [Mpc/h]    & Number of particles   \\ \toprule
\textsc{CosmoGridV1}    & $\Omega_m$, $\sigma_8$, $H_0$, $w_0$, $n_s$, $\Omega_b$               & 2500\x7                                            & \centering 200                                     & \centering 900      & 832$^3$               \\ 
\textsc{cosmo-SLICS}    & $\Omega_m$, $\sigma_8$, $H_0$, $w_0$                                  & 25\x2                                              & \centering 800                                     & \centering 505      & 1536$^3$              \\ 
\textsc{DarkGridV1}     & $\Omega_m$, $\sigma_8$                                                & 58\x5                                              & \centering 50                                      & \centering 900      & 768$^3$               \\ 
\textsc{MassiveNuS}     & $\Omega_m$, $\sigma_8$, $M_\nu$                                       & 101\x1                                             & \centering n/a                                     & \centering 512      & 1024$^3$              \\ 
\textsc{DH10}           & $\Omega_m$, $\sigma_8$                                                & 158\x1                                             & \centering n/a                                     & \centering 140      & 256$^3$               \\\bottomrule
\hline
\end{tabular}
}
\caption{
List of simulation sets used in map-level simulation-based inference of cosmological parameters from LSS probes, either for forecasts or measurements. 
The \textsc{cosmo-SLICS} simulations description is taken from \citet{HarnoisDeraps2019cosmoslics}, the \textsc{MassiveNus} from \citet{Liu2018massiveNus}, the \textsc{DarkGridV1}, a precursor to \cosmogrid, from \citet{Zuercher2022despeaks}, and the DH10 from \citet{DietrichHartlap2010peaks}.
This list includes only simulations that have lightcone shells/snapshots output dense enough to enable map-level inference with projected probe maps from LSS surveys.
This list is not exhaustive; other notable datasets were used by \citet{liu2014cosmology,ZorrillaMatilla2016,Fluri2018deep}.
}
\label{tab:sims_compare}
\end{center}
\end{table*}

\section{The cosmological parameter grid} \label{sec:grid}

\begin{figure*}
\centering
\includegraphics[width=0.9\textwidth]{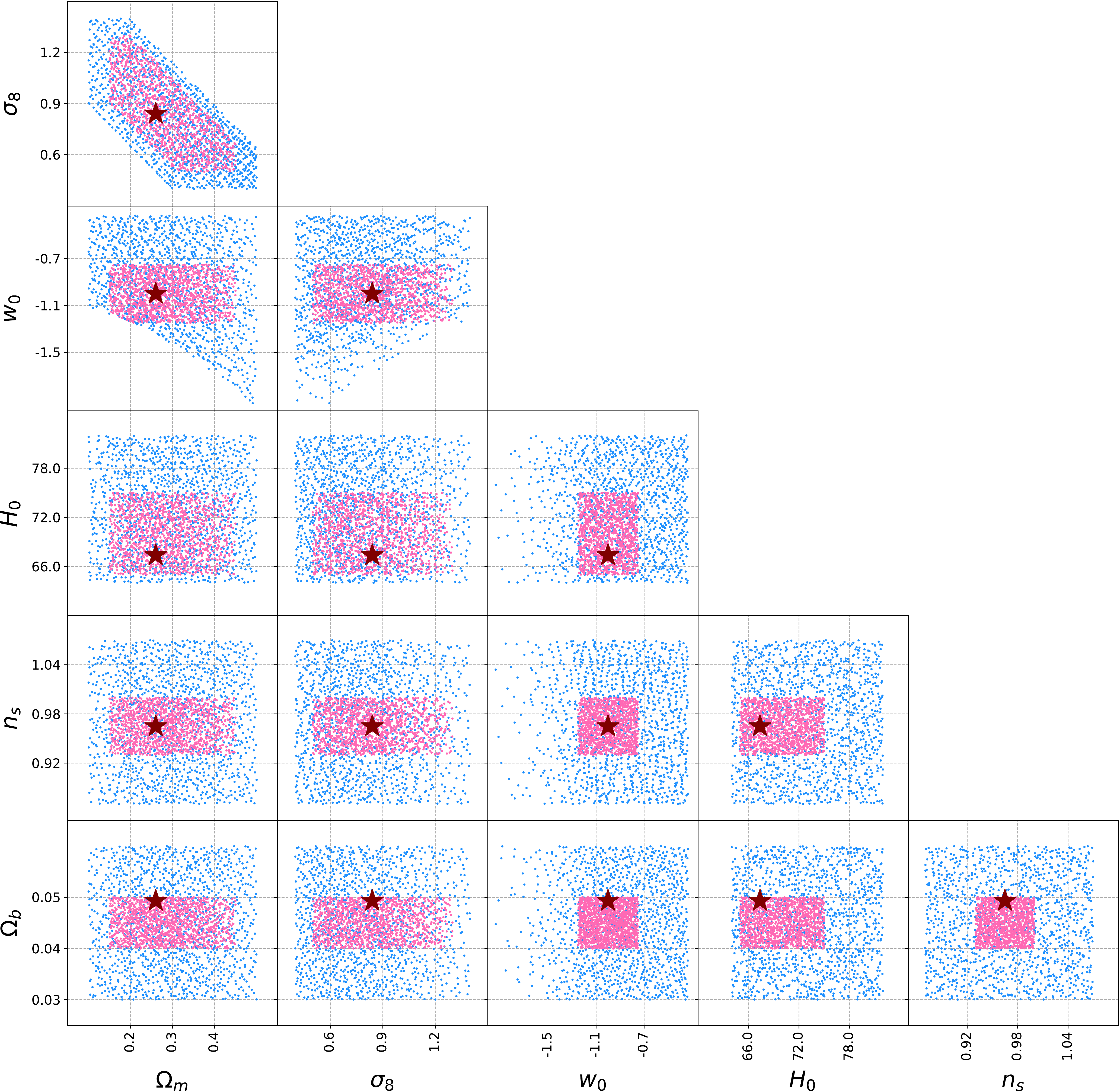}
\caption{2500 cosmological parameters used in \cosmogrid. Each grid point has 7 N-body runs from unique initial conditions. The blue and pink parameters belong to the ``wide'' and ``narrow'' grids, respectively. The fiducial cosmology is marked with a red star and contains 200 unique runs.}
\label{fig:points}
\end{figure*}

The \cosmogrid\ parameter points were chosen to span the \wcdm\ model and include the following parameters: matter density today \omatter, matter clustering amplitude \sigeight, the Hubble constant $H_0$, the dark energy equation of state \wzero, the spectral index \ns, and the baryon density today \obary.
We do not vary the neutrino mass, and we fix it to to three degenerate neutrinos, each with $m_\nu$ = 0.02 eV. 

The priors were chosen to be uniformly spaced in the parameters listed above. 
There is another additional restriction on the prior in the \omatter -- \wzero\ plane, which removes some combinations of these parameters from the simulation set. 
This is due the effective nature of the \wcdm\ model for modeling of the relativistic fields, which includes neutrinos, and is in nature similar to the phantom crossing for dark energy models.
To model this, we used the method available in \pkdgrav, based on \citet{Tram2019neutrinos}.
In this method, the inclusion of relativistic fields requires computing the transfer functions in the N-body gauge.
We generate the transfer functions with the \textsc{Class} code \citep{Lesgourgues2011class}.
Then, we transform them to the N-body gauge using the \concept\ code \citep{Dakin2019concept}.
To do this, we need to calculate the total velocity transfer function $\vartheta_{\rm{tot}} = \sum_i \vartheta_i$, which sums over the all relativistic species~$i$.
Each term is dependent on its density and pressure $(\rho + p)_i$, and therefore the normalization of the $(\rho + p)_{\rm{tot}}$ may become negative if $w<-1$.
This can happen at different times in cosmic history for different cosmological models, and depends most strongly on \omatter.
It is therefore possible to run simulations for some set of models, for which that transition does not happen before $z$=0.
This prevents us from running simulations at some models for $w<-1.1$.
See Appendix~A in \citetalias{Fluri2022kids} for more details about the nature of this model.
We also restricted the \omatter -- \sigeight\ prior to lie along the \Seight\ degeneracy, removing the corner combinations.
These combinations are excluded according to observations.

The variable parameters were sampled on a 6-dimensional Sobol sequence \citep{Sobol1967sequence}. 
A Sobol sequence is a deterministic low-discrepancy sampling scheme, with the property that it can be easily extended in both the number of samples and number of dimensions.
This makes it easy to add new parameters, such as baryonic feedback or intrinsic alignments (see sections~\ref{sec:baryons} and~\ref{sec:intrinsic_alignments}).
Futhermore, the grid is divided into the \emph{wide} and \emph{narrow} priors, split evenly;
we create more simulations in the area suggested by other observations, such as the CMB, while still maintaining the capacity to use wide priors.
This way we allow for the results of \cosmogrid\ -based inference to be compared with main survey analysis, which typically uses broad, uninformative priors \citep{Des2022combined}.
\rev{The points in the narrow grid are chosen from the same sequence as the wide: we continued sampling the wide sequence and simply discarded all points outside the narrow grid box.}

The 200 fiducial simulations are taken at the fixed cosmology, each with a different seed for the initial conditions.
The large number of independent simulations allows for computation of precise covariance matrices for the chosen features, if the covariance computation is part of the inference scheme employed, such as it was used by \citetalias{Zuercher2022despeaks}, for example.
For each variable parameter, we also simulate two $\pm \Delta$ runs, with the same initial conditions as the fiducial run.
This enables the calculation of the stencil derivatives of the features for each parameter, which can be useful in different inference schemes \citep{Charnock2018imnn,Fluri2021inference}.
The fiducial cosmology parameters were chosen to match previous CMB observations \citep{Planck2018parameters}, with values and derivative steps listed in Table~\ref{tab:prior}.

\begin{table}
\begin{center}
\begin{tabular}{ l l l l l}
          & fiducial &$\Delta$ fid. & wide grid prior           & narrow grid prior        \\ \hline
\omatter  & 0.26     &$\pm$ 0.01    & $\in$ [ 0.10, \ 0.50 ]    & $\in$ [ 0.15, \ 0.45 ]   \\
\sigeight & 0.84     &$\pm$ 0.015   & $\in$ [ 0.40, \ 1.40 ]    & $\in$ [ 0.50, \  1.30 ]  \\
\wzero    & -1       &$\pm$ 0.05    & $\in$ [-2.00, -0.33 ]     & $\in$ [-1.25, -0.75 ]    \\
\ns       & 0.9649   &$\pm$ 0.02    & $\in$ [ 0.87, \ 1.07 ]    & $\in$ [ 0.93, \ 1.00 ]   \\
\obary    & 0.0493   &$\pm$ 0.001   & $\in$ [ 0.03, \ 0.06 ]    & $\in$ [ 0.04, \ 0.05 ]   \\ 
\hubble   & 67.3     &$\pm$ 2.0    & $\in$ [ 64.0, \ 82.0 ]     & $\in$ [ 65.0,  \ 75.0 ]  \\ \hline
\end{tabular}
\caption{Cosmological parameters in \cosmogrid. Parameters \sigeight, \omatter, \wzero have additional restrictions beyond the box prior listed here, as described in Section~\ref{sec:grid}. }
\label{tab:prior}
\end{center}
\end{table}

The simulation grid points are shown in Figure~\ref{fig:points}. The ``wide'' and ``narrow'' priors are shown in blue and pink, respectively. 
The fiducial cosmology is marked with a red star.
The boundaries of the prior for the cosmological parameters is shown in Table~\ref{tab:prior}.

The benchmark simulations use the same cosmological parameters as the fiducial. 
We created three benchmarks that enable us to test the dependence of analysis choices on the following simulation parameters: (i) increased box size, (ii) increased number of particles, (iii) increased redshift resolution of shells.
These choices are described in the following Section~\ref{sec:sims}.

\section{Dark matter simulations} \label{sec:sims}

Choosing the configuration of box size and particle count that is sufficient for a given type of survey analysis requires considering multiple trade-offs, as described in Section~\ref{sec:theory_prediction}.
In \cosmogrid, we targeted Stage-III LSS surveys, with the focus mainly on the DES and KiDS datasets, as well as intermediate length-scales, of \mbox{$\ell<1500$}.
The target effective galaxy number density was $n_{\rm{eff}}=10$ galaxies per arcmin$^2$ and redshifts $z\lesssim1.5$.
Below we describe our choices regarding the box size, number of particles, and shell spacing.
The summary of the \cosmogrid\ simulation parameters is presented in Table~\ref{tab:sim_params}.

\paragraph{Box size.} Using projections of the same matter density distribution rotated by different angles is a common practice in creating lightones, whether with a pencil-beam approach, or via box replication. 
For sufficiently large boxes, this technique enables us to increase the variance in the projected maps distribution and bring it closer to the true cosmic variance.
For a box replication lightcone approach, cutting out survey maps from different regions of the sky corresponds to rotating the replicated boxes \citepalias{Zuercher2022despeaks,Fluri2022kids}. 
For the pencil-beam approach, re-using the same volume can lead to discontinuities in lightcone density, which will affect a small area of the total volume.
These discontinuities can introduce unphysical features and give raise to biases in parameter measurements. 
For the box replication scheme, which employs periodic boundary conditions, there are no discontinuities along the box boundary, but the structures are replicated, which can affect the variance of the features chosen.
\citet{Sgier2021combined,Matilla2020optimizing} explored these biases for \cl\ and other statistics.
The effects of box replication can be tested using the benchmark simulations with bigger boxes (see Section~\ref{sec:benchmarks}).

In \cosmogrid\ we follow the box replication approach, aiming at having no more than roughly 3-4 replicas inside a line of sight up to $z$=1.5, to avoid cosmic variance effects.
However, in the replicated lightcone simulation, some lines of sight along the box edges will still go through the same structures.
Although their density will have evolved throughout cosmic time and most likely had enough time to decorrelate, we employ the \emph{shell permutation scheme} to remove them completely;
we construct lightcones from up to 7 independent simulations (see Section~\ref{sec:lightcone_permutations}).
This led to the choice of medium-sized boxes \lbox=900 Mpc/h.
To test the validity of these assumptions, we also create a benchmark with box size \lbox=2250 Mpc/h.
The methodology for testing analysis choices against the benchmarks in described in Section~\ref{sec:benchmarks}.

\paragraph{Particle count.} To choose the number of particles, we consider the shot noise contribution to the maps due to the particle count.
We calculate that, for boxes of around \mbox{\lbox=1 Gpc/h}, \npart=1024$^3$, and the lensing maps created with 10 galaxies/arcmin$^2$, the shot noise contribution to the \cl\ reaches 5\% at $\ell \sim 1500$. 
That makes the shot noise resulting from particle counts roughly 200\x\ smaller than the shape noise from weak lensing at this galaxy density.
As high particle count leads to increased runtime, we choose of the number of particles to be \npart=832$^3$.
To be able to further test this, we create benchmark simulations with \npart=2048$^3$ particles and the same box size as the main simulations.

\paragraph{Redshift resolution.} We provide Healpix maps of particle counts for a fixed number of shells dividing the full particle lightcone. 
Shell thickness is chosen with respect to the precision requirements for projected maps and available hard drive storage.
The required shell thickness was studied by \citet{Matilla2020optimizing}, who found thickness of $\sim$ 60 Mpc/h to be  sufficient. 
We decide to store the maps with $\sim 70$ shells in the redshift range $z\in[0,3.5]$, with shell thickness increasing with redshift.
Due to the fact that \pkdgrav\ takes steps in proper time, which is cosmology dependent, the shell boundaries are slightly different for each cosmology.
The final number of shells is 69, with the exception of 6 models that have 68.
In this configuration for the fiducial cosmology, shells at $z=0, 0.5, 1.5$ have the corresponding thickness of $\Delta_{\rm{shell}} \approx 40,55,90$ Mpc/h.
We run \pkdgrav\ with 140 base timesteps from $z$=99, with the first 70 steps equally spaced in proper time evolving the particles up to $z$=4.
The remaining 70 steps were also spaced equally in proper time between $z$=4 and $z$=0, and produce shell output in lightcone mode.
\rev{The shell boundaries in redshift are stored for each simulation, so they can be used later during the shell projection steps.}

\begin{table}
\begin{center}
\setlength{\tabcolsep}{2.5pt}
\footnotesize{\def\arraystretch{1.3}
\begin{tabular}{
 >{\raggedright\arraybackslash} p{1.4cm} 
 >{\raggedright\arraybackslash} p{1.75cm}  
 >{\raggedright\arraybackslash} p{2.2cm} 
 >{\raggedright\arraybackslash} p{1.6cm} 
 >{\raggedright\arraybackslash} p{1.6cm} 
 >{\raggedright\arraybackslash} p{1.3cm} 
 >{\raggedright\arraybackslash} p{1.25cm} 
 >{\raggedright\arraybackslash} p{2.9cm}
}
\textsc{CosmoGridV1} \\
\toprule
Set                             & Number of cosmologies     & Number of simulations per cosmology & Box size $L_{\rm{box}}$ [Mpc/h] & Number of particles $n_{\rm{part}}$ &  Number of base timesteps &  Dataset size [TB] & Comments \\[3em] \toprule
Fiducial                        & \centering 1+2$\times$6   & \centering 200                      & \centering 900                  & \centering 832$^3$                  & \centering 140            & \centering 14      & fiducial cosmology and central difference derivatives \\
Grid                            & \centering 2500           & \centering 7                        & \centering 900                  & \centering 832$^3$                  & \centering 140            & \centering 96      & 6-D Sobol sequence spanning $w$CDM \\ \midrule
Benchmark baseline              & \centering 1              & \centering 7                        & \centering 900                  & \centering 832$^3$                  & \centering 140            & \centering 0.02    & Baseline simulations for benchmarks, same as fiducial \\
Benchmark particle count        & \centering 1              & \centering 7                        & \centering 900                  & \centering 2048$^3$                 & \centering 140            & \centering 0.2     & convergence test for simulation resolution \\
Benchmark box size              & \centering 1              & \centering 7                        & \centering 2250                 & \centering 2080$^3$                 & \centering 140            & \centering 0.11    & convergence test for box size, same particle density as the fiducial \\
Benchmark redshift resolution   & \centering 1              & \centering 7                        & \centering 900                  & \centering 832$^3$                  & \centering 500            & \centering 0.42    & convergence test for number of shells per simulation  \\\bottomrule
\hline
\end{tabular}

}
\caption{Summary of \cosmogrid\ simulation parameters for the grid, fiducial and benchmarks sets.}
\label{tab:sim_params}
\end{center}
\end{table}

For the main configuration, the resulting particle mass for the fiducial cosmology is $M=9.08\x10^{10} \ h^{-1} M_\odot $. 
The particle mass is cosmology dependent and ranges from $M=3.5 - 17.5\x10^{10} \ h^{-1} M_\odot$
The softening length is kept at the fiducial \pkdgrav\ value of 0.02\x\ mean particle separation.
\rev{The opening angle $\theta$, which controls the force accuracy calculation, was kept at \pkdgrav\ defaults.}
For the halo finder, the linking length used 20\% of the mean particle separation.

\begin{figure*}
\centering
\includegraphics[trim={0 0.8cm     0.15cm 0},   clip, width=0.355\textwidth]{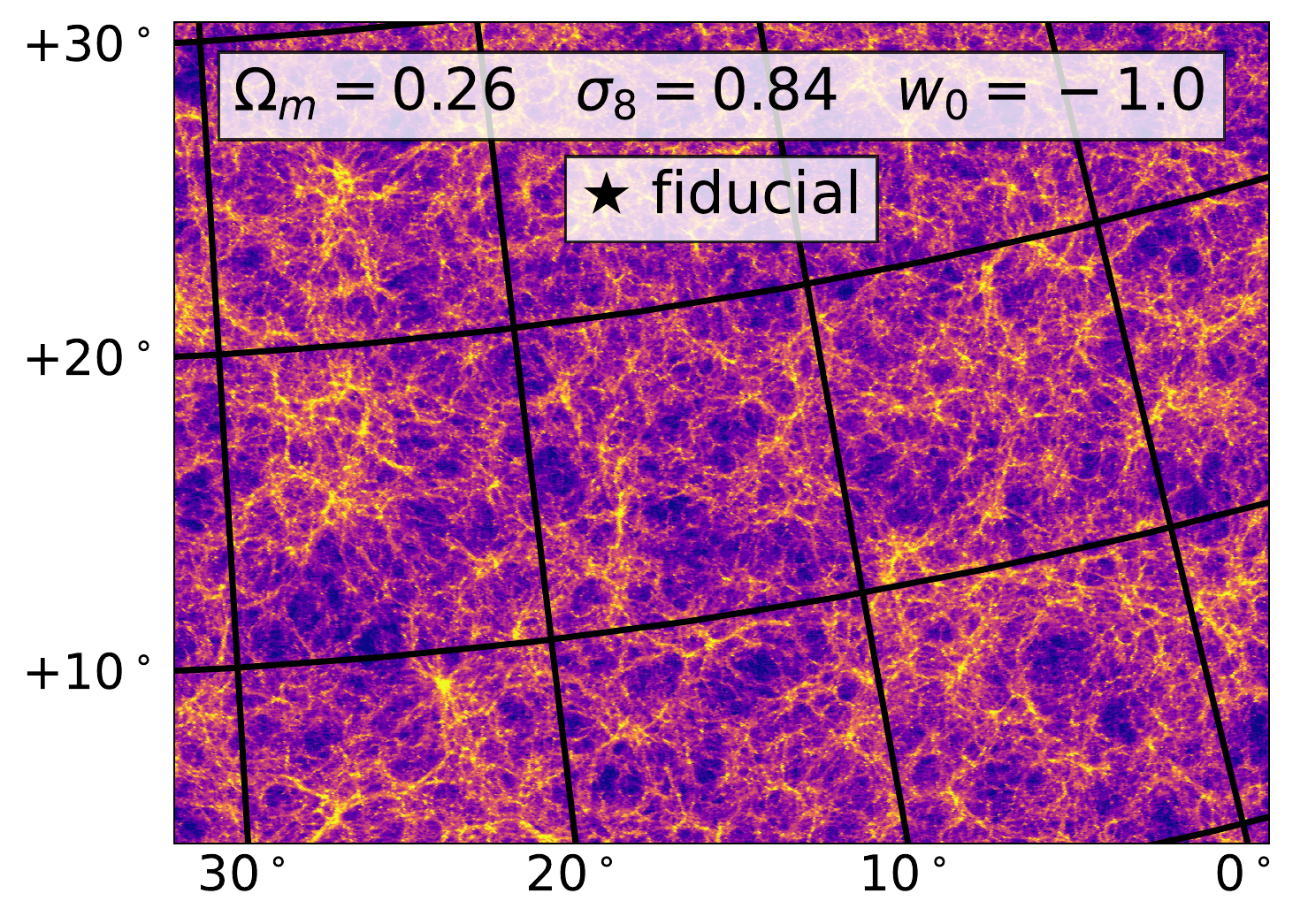}
\includegraphics[trim={1.9cm 0.8cm 0.15cm 0},   clip, width=0.31\textwidth]{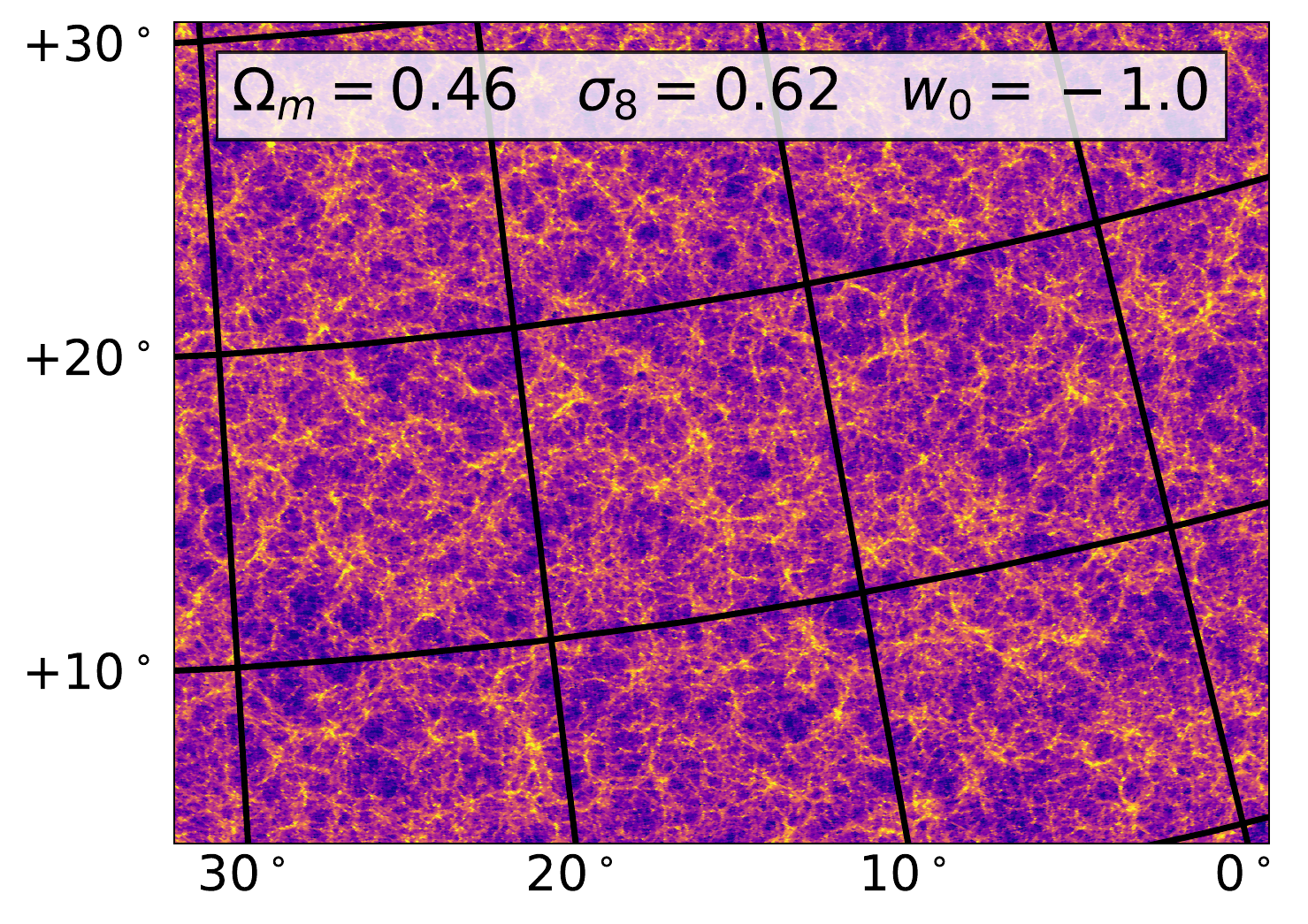}
\includegraphics[trim={1.9cm 0.8cm 0.15cm 0},   clip, width=0.31\textwidth]{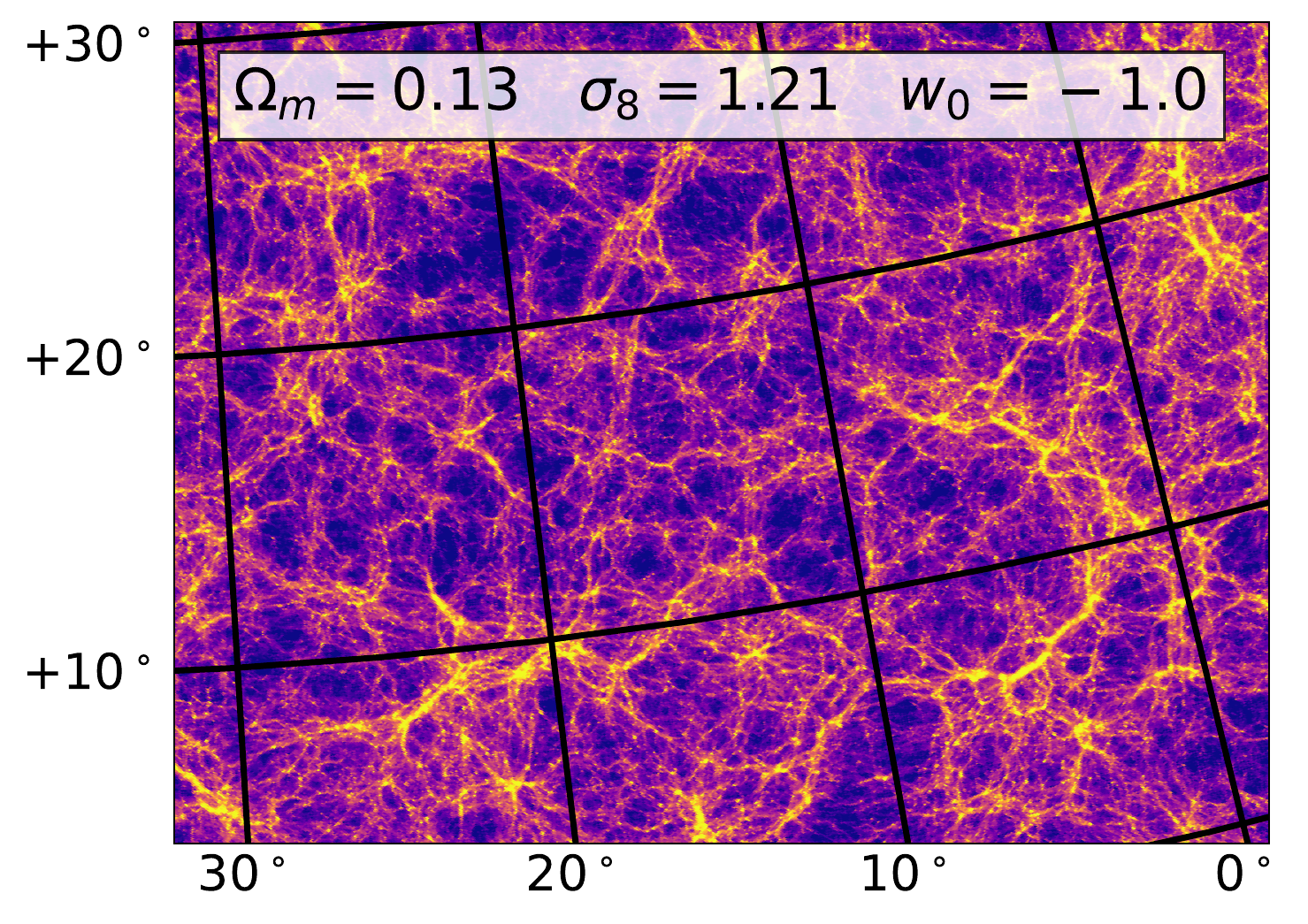}
\includegraphics[trim={0 0         0.15cm 0},   clip, width=0.355\textwidth]{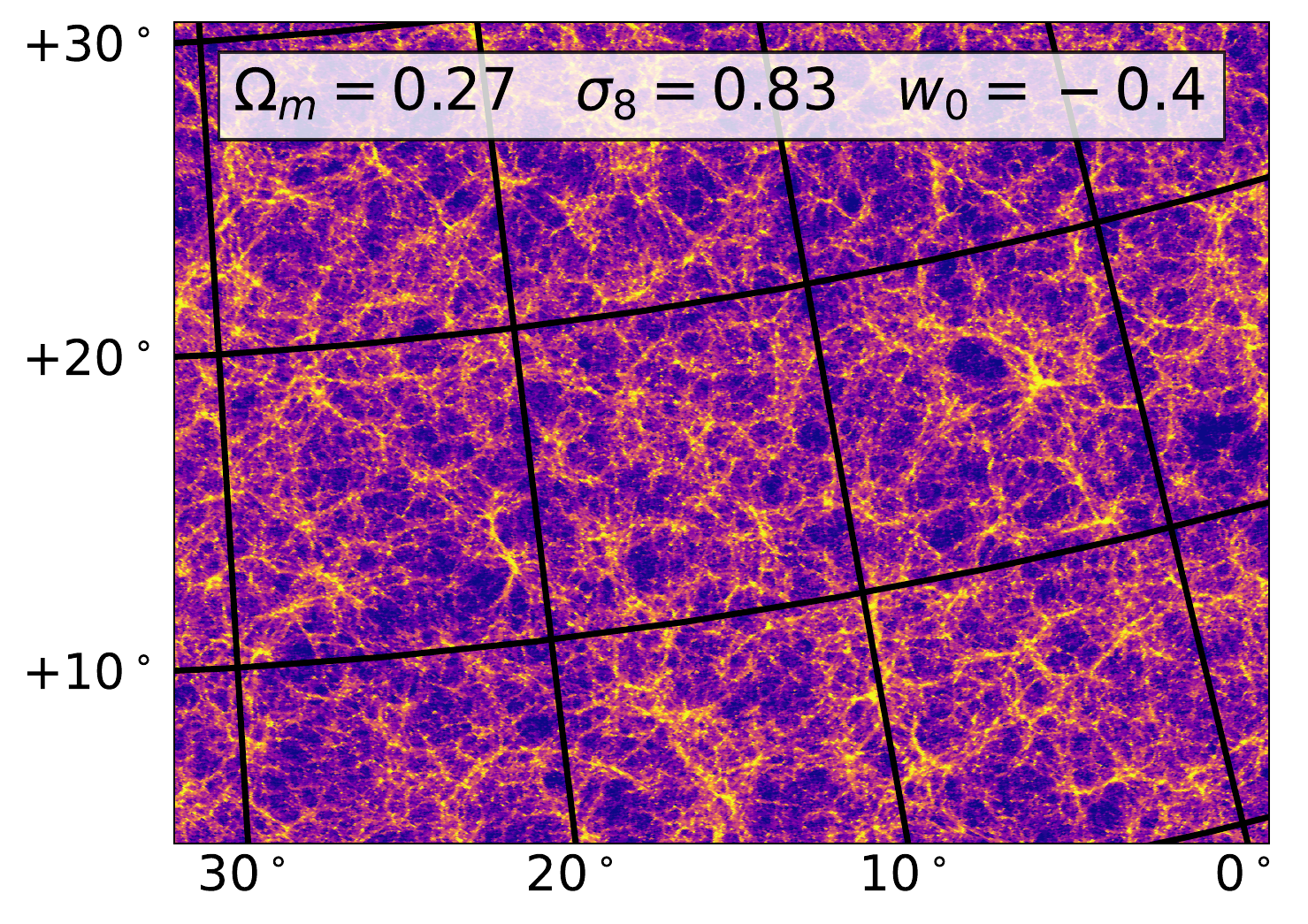}
\includegraphics[trim={1.9cm 0     0.15cm 0},   clip, width=0.31\textwidth]{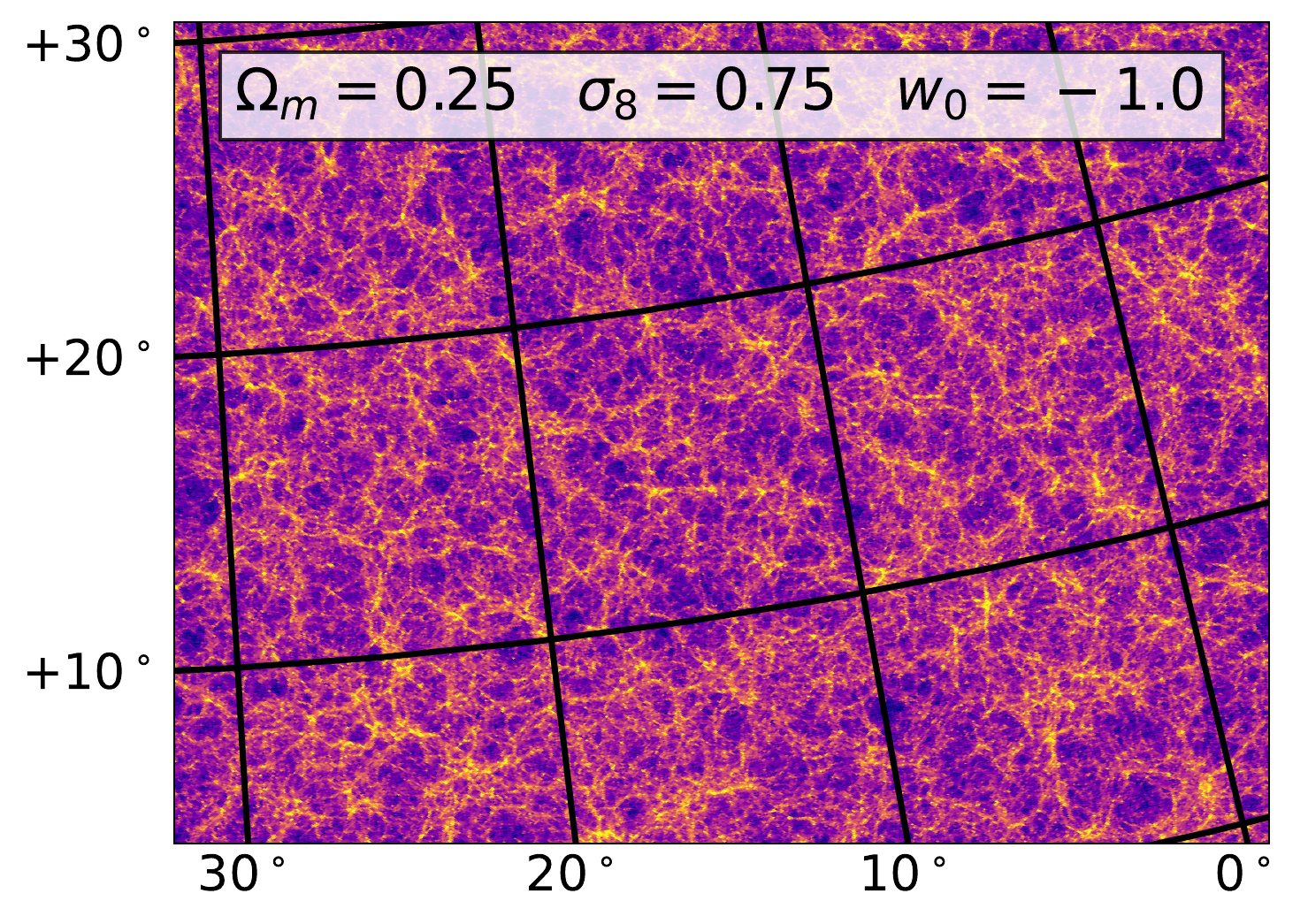}
\includegraphics[trim={1.9cm 0     0.15cm 0},   clip, width=0.31\textwidth]{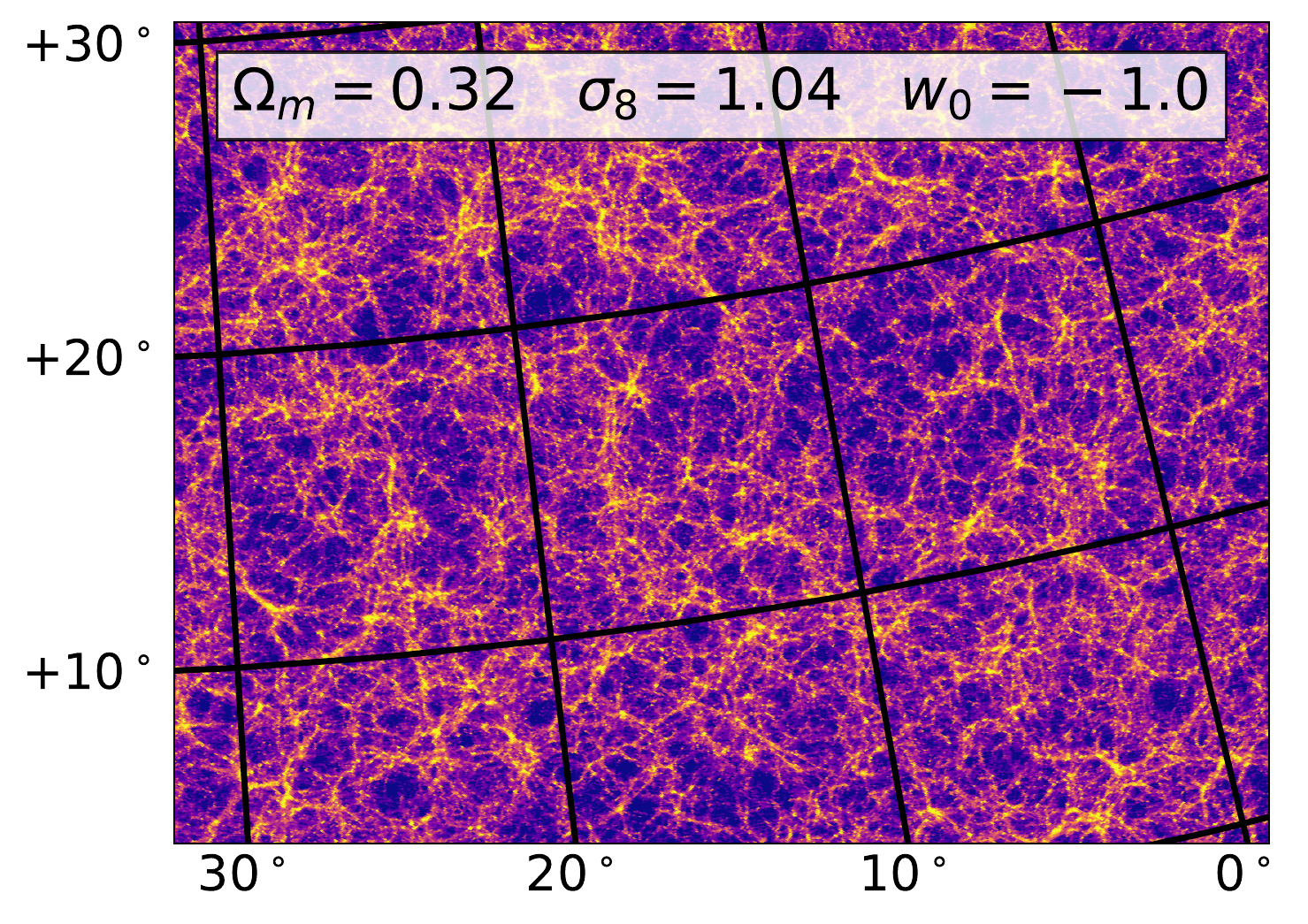}
\caption{
Example spherical simulation shells from the fiducial cosmological model (top left) and other models in the grid. 
The maps show the logarithm of the density contrast $\delta = (\rho-\bar \rho)/(\bar \rho)$. 
All maps come from different initial conditions. 
The top/bottom middle and right panels show extreme cosmologies along/across the \Seight\ degeneracy. 
The bottom left panel shows a cosmology close to the fiducial, but with an extreme \wzero. 
The dynamic range of the plotted density constrast is different for each cosmology, and spans the range between 2-98 percentile.
}
\label{fig:example_shells}
\end{figure*}

\begin{figure}
\centering
\includegraphics[width=0.95\textwidth]{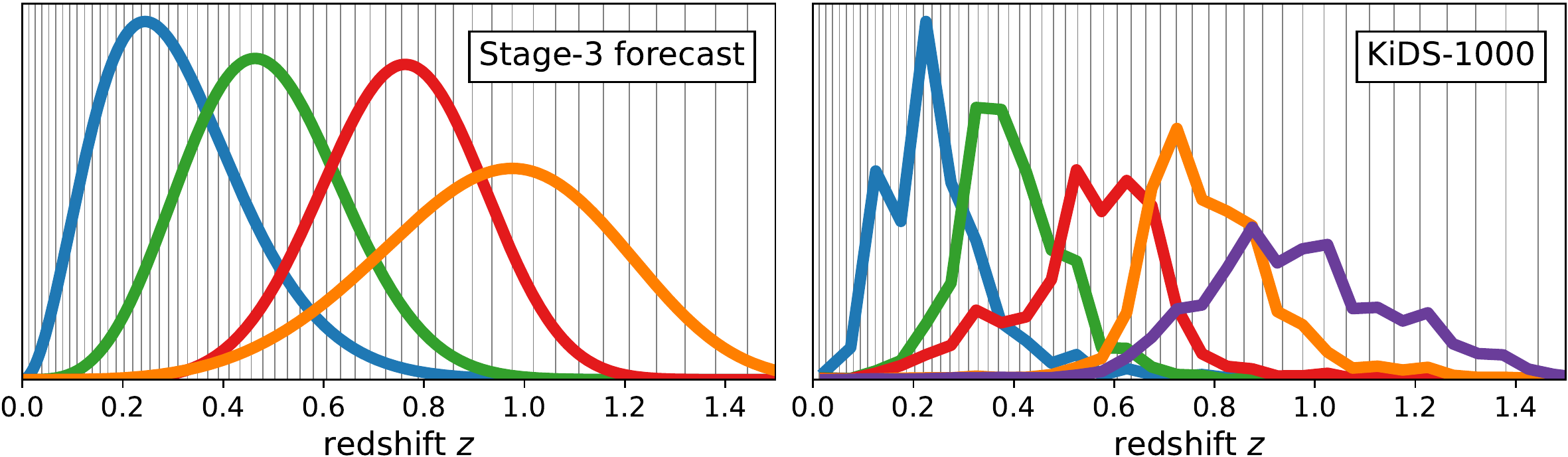}
\caption{
Shell boundaries of the fiducial cosmology (grey vertical lines) and tomographic redshift distribution of weak lensing sources $n(z)$ for a Stage-III forecast \citep{Fischbacher2022ia} and KiDS-1000 \citep{hildebrandt2021kids}, as used in \citetalias{Fluri2022kids}.
Projected maps with these kernels are available as a part of the \cosmogrid\ public data release.
}
\label{fig:shells_nz}
\end{figure}

Figure~\ref{fig:shells_nz} shows the shell boundaries as a function of redshift compared to $n(z)$ of weak lensing source galaxies for two datasets: KiDS-1000 \citep{hildebrandt2021kids} and Stage-III forecast from \citet{Fischbacher2022ia}, which were modeled on DES-Y3 \citep{myles2021dark}.
We show the shell boundaries for the fiducial cosmology up to $z$=1.5.
The shells are stored at Healpix \nside=2048 and contain the number count of particles inside a pixel.
We create a set benchmark simulations with the number of base timesteps in \pkdgrav\ increased to 500, with 100 step down to $z$=4 and 400 steps from $z$=4 to $z$=0.

The example spherical maps for different cosmological models are shown in Figure~\ref{fig:example_shells}.
The points were chosen to illustrate the differences between the fiducial model (upper left) and the grid:
Top middle and top right panels shows extreme models \emph{along} the \Seight\ degeneracy, while the lower middle and right panels shows models \emph{across} that degeneracy.
The lower left panel shows a cosmology similar to the fiducial, but with extreme dark energy equation of state \wzero.
The maps show the log density contrast $\delta = (\rho-\bar \rho)/(\bar \rho)$.
The colorscale is chosen from the fiducial cosmology, spanning the dynamical range between the 2nd and 98th percentile of the sphere full map.
The maps shown here are at the resolution of \nside=1024 at $z\approx0.5$.
The panels clearly show how different the maps of large scale structure are for varying cosmological models. 
These complex pattern differences can be exploited by non-Gaussian statistics and machine learning.

\section{Observable maps projection} \label{sec:maps}

To create forward-modeled maps of observable probes, we project the lightcone shells against probe kernels corresponding to the tomographic galaxy samples.
We follow the formalism used in \citet{Sgier2018,Sgier2021combined}, where the Born approximation was used with the \ufalcon\ code\footnote{\url{https://cosmology.ethz.ch/research/software-lab/UFalcon.html}}.
The Born approximation was found to be sufficiently precise for lensing analysis using simulations for intermediate scales by \citet{Petri2017born,Fluri2019kids}.
We create the projected maps $m_{\mathrm{2D}}$ of lensing convergence, galaxy density, and intrinsic alignment in the following way:
\begin{align}
\label{eqn:projection}
m_{\mathrm{2D}}^{\mathrm{pix}} &\approx \sum_bW^{m}\int_{\Delta z_b}\frac{\mathrm{d}z}{E(z)}\delta_{\mathrm{3D}}\left[\frac{c}{H_0}\mathcal{D}(z)\hat{n}^\mathrm{pix},z\right],
\end{align}
where $W^m$ is the relevant probe kernel, 
$\hat{n}^{\mathrm{pix}}$ is a unit vector pointing to the pixels center,
$\mathcal{D}(z)$ is the dimensionless comoving distance,
$E(z)$ is given by $\mathrm{d}\mathcal{D} = \mathrm{d}z/E(z)$,
and $\Delta z_b$ is the thickness of shell $b$.
The kernels $W$ for weak lensing (WL), intrinsic alignments (IA), and galaxy clustering (G) are calculated from the map shells following \citet{KacprzakFluri2022deeplss}:
\begin{align}
\label{eqn:projection_kernels}
W^{\mathrm{WL}} & = \frac{3}{2}\Omega_\mathrm{m}  \frac{\int_{\Delta z_b}\frac{\mathrm{d}z}{E(z)}\int_z^{z_s}\mathrm{d}z'n(z')\frac{\mathcal{D}(z)\mathcal{D}(z,z')}{\mathcal{D}(z')}\frac{1}{a(z)}}{\int_{\Delta z_b}\frac{\mathrm{d}z}{E(z)}\int_{z_0}^{z_s}\mathrm{d}z'n(z')} \\[1em]
W^\mathrm{IA} &= \frac{\int_{\Delta z_b}\mathrm{d}zF(z)n(z)}{\int_{\Delta z_b} \frac{\mathrm{d}z}{E(z)}\int_{z_0}^{z_s}\mathrm{d}z'n(z')} \\[1em]
W^\mathrm{G} &= \frac{\int_{\Delta z_b}\mathrm{d}z \ n(z)}{\int_{\Delta z_b} \frac{\mathrm{d}z}{E(z)}\int_{z_0}^{z_s}\mathrm{d}z'n(z')} 
\end{align}
where $n(z)$ is the redshift distribution of galaxies, $z_s$ and $z_0$ are the source and observer redshifts, respectively, and $F(z)$ is a cosmology and redshift dependent term:
\begin{equation}
F(z) = -C_1\rho_\mathrm{crit}\frac{\Omega_\mathrm{m}}{D_+(z)},
\end{equation}
where ${C_1 = 5 \times 10^{-14} \ h^{-2}M_\odot\mathrm{Mpc}^3}$ is a normalization constant, $\rho_\mathrm{crit}$ is the critical density at $z$=0 and $D_+(z)$ normalized linear growth factor, so that $D_+(0) = 1$. 

\begin{figure*}
\centering
\includegraphics[width=1\textwidth]{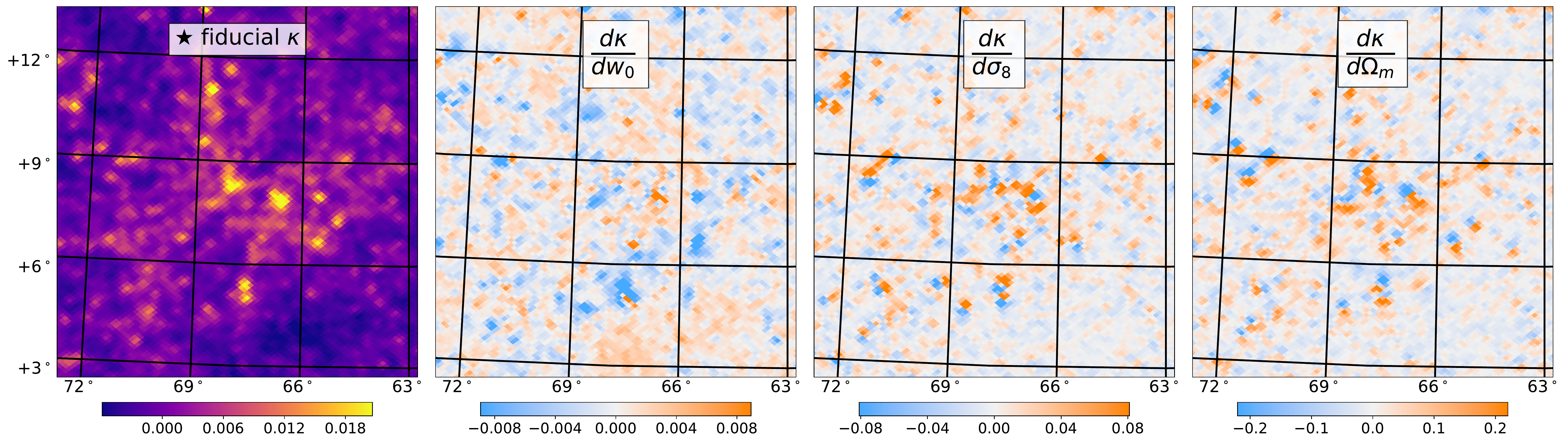}
\caption{
An example projected map from the fiducial cosmology (left) and its derivatives with respect to the \wzero, \omatter, and \sigeight\ parameters. 
The derivative maps are created from the $\pm\Delta$ simulations described in Section~\ref{sec:grid}.
}
\label{fig:example_deltas_maps}
\end{figure*}

An example lensing convergence map for the second redshift bin in the Stage-III set is shown on the left panel of Figure~\ref{fig:example_deltas_maps}.
This map was taken from the fiducial cosmological model. 
The other panels show the derivative of this map with respect to three cosmological parameters: \wzero, \omatter, and \sigeight.
As expected, changes in \wzero\ mostly affect the depth of voids and height of peaks, while changes in \omatter\ and \sigeight\ also modify the configuration of positions of the halos.

\rev{Full ray-tracing is likely to be needed for a high resolution analysis. 
An implementation of ray-tracing should be possible during the projection step for a cost of increased computing time.
We leave this to future work.}

\subsection{Shell permutations} \label{sec:lightcone_permutations}

Using box replication allows to maintain the balance between particle density and box size within the feasible computational limits, but risks introducing errors in the variance of the maps. 
As pencil-beam lightcones are often created with multiple simulation boxes rotated by a random angle.
In a box replication mode, the replicas are not rotated; this results in some line of sight angles being integrated with the same structures at different redshifts.
These lines of sight are at multiples of 45$^\circ$.
The lines of sight at multiples of 45$^\circ/n$ will go through the same point in the box every $n$-th replica. 
Other lines of sight will view the neighboring boxes at different angles, which is similar to the pencil-beam approach.
As 45$^\circ$ lines of sight are rare, it will not introduce significant biases into cosmological measurements, especially for probes that integrate over broad kernels, such as weak lensing.

To eliminate this effect completely, we created a novel \emph{shell permutation} scheme.
In this scheme, we divide the lightcone into groups of shells, each group taken from an independent simulation with different initial conditions.
The groups of shells are chosen such that the outer $z$-border of the last shell in the group is crossing the size of the box, which in our case is 900 Mpc/h.
The lightcone up to $z$=3.5 typically consists of around 6 replicated boxes, although that can vary with cosmology. 
As \cosmogrid\ has 7 independent simulations per grid point, this allows us to create lightcones with all shell groups coming from unique initial conditions.
This way no line of sight will go through the same points in the simulation volume across multiple replicas.
Additionally, we randomly flip (up-down and left-right) and rotate (4 possible Healpix symmetries) each shell group before projecting them into probe maps.
This allows to increase the number of semi-independent realizations of the maps even further, bringing the variance of the maps closer to the true cosmic variance.
For the pencil-beam scheme, a similar approach was developed by \cite{Giocoli2018dustgrain}.

\begin{figure*}
\centering
\includegraphics[width=0.8\textwidth]{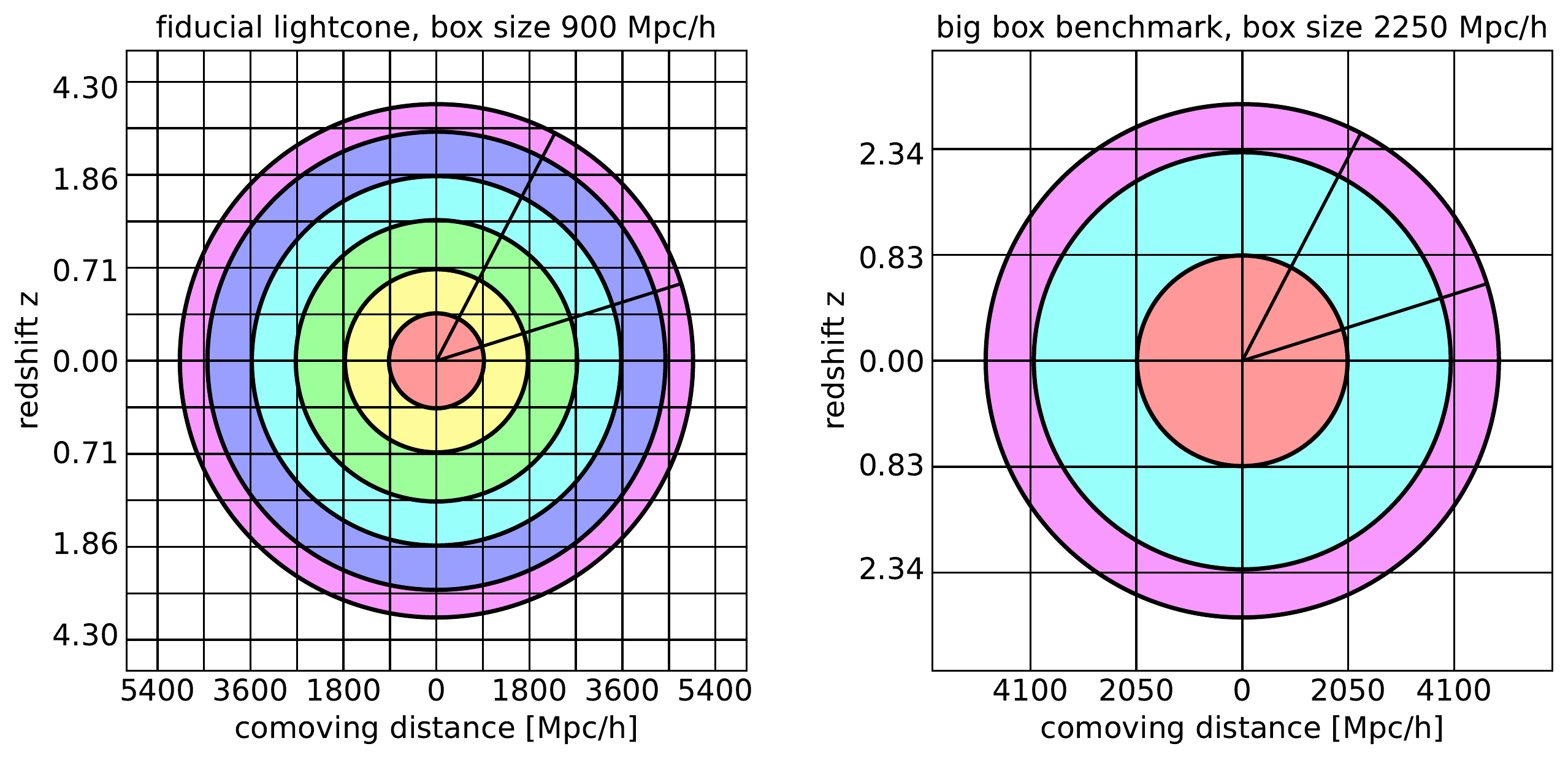}
\caption{
Illustration of the box replication and shell permutation scheme used in \cosmogrid. 
The square grid shows the periodic boundaries of the simulation boxes, for the fiducial cosmology.
Lightcone volume covered by shell groups taken from separate simulations with unique initial conditions are shown with different colors.
The left panel shows the main configuration used in the grid and fiducial runs, and the right shows the big box benchmark.
The opening angle for a 5000 deg$^2$ survey is shown.
}
\label{fig:box_replication}
\end{figure*}

The lightcone construction with shell permutations is shown in Figure~\ref{fig:box_replication}, for the main simulation configuration and for the big box benchmark. 
The shell groups, each coming from an independent simulation, are shown using the colored rigs.
Using shell permutations introduces additional discontinuity areas into the lightcone, but we find it to be a negligible fraction.
We calculated that, for voxels of size 4 Mpc/h, the fraction of volume affected by discontinuities from this scheme is around 0.1\%.

\section{Baryonic feedback} \label{sec:baryons}

Small scales in the matter density distribution are likely to be affected by baryon feedback effects \citep[see][and refrences therein]{Chisari2019baryons}, which include the subgrid physical effects created by stellar and Active Galactic Nuclei (AGN) activity.
The uncertainty of effects of baryons is likely to increase with decreasing length scales.
These are simulated using hydrodynamic codes, which are typically very expensive to run.
A faster way to include baryons is to use effective models that can transform the dark matter only density distribution to a corresponding distribution that would arise if baryons were fully included.
One such model is called \emph{baryonification} and was proposed by \citet{Schneider2019baryon}.
There, the baryon feedback effects were introduced by modifying the particle positions inside a dark matter-only N-body simulation snapshot. 
It was found to induce effects that span the space covered by hydro-simulations.
An advantage of baryonification is that it can be applied relatively fast in post-processing of dark matter only simulations, for different baryon feedback parameter models.
Priors on these parameters can be obtained by external observations.

\subsection{The baryonification model}

The baryonification method describes the density field as a set of halos with profiles affected by dark matter 1-halo and 2-halo components.
The dark matter only (dmo) field is described as a set of halos with density 
\begin{equation}
    \rho_{\mathrm{dmo}}(r) = \rho_{\mathrm{NFW}}(r) + \rho_{\mathrm{2h}}(r),
\end{equation}
where $\rho_{\mathrm{NFW}}(r)$ corresponds to a generalized NFW profile, which depends on the cosmological parameters of the simulations, the viral mass of the halo $M \equiv M\vir$ and the concentration \mbox{$c \equiv c\vir$}, enclosed within the radius in which the density is 200\x\ larger than the average density.
Within the baryonification model (dmb), the 1-halo term contains the collisionless components (clm), the gas component (gas) and the central galaxy (cga).
The $\rho_{\mathrm{clm}}$ is dominated by dark matter but also contains satellite galaxies and intracluster stars.
The total baryonified profile $\rho_{\mathrm{dmb}}(r)$ is:
\begin{equation}
    \rho_{\mathrm{dmb}}(r) = \rho_{\mathrm{clm}}(r) + \rho_{\mathrm{gas}}(r) + \rho_{\mathrm{cga}}(r) + \rho_{\mathrm{2h}}(r).
\end{equation}
The baronification model transforms the simulation with density $\rho_{\mathrm{dmo}}$ into $\rho_{\mathrm{dmb}}$.
This is done using the integrated mass profile function inside radius $r$, defined as
\begin{equation}
    M_\chi(r)=\int_0^r s^2 \rho_\chi (s) \mathrm{d} s
\end{equation}
which is a bijective function and can be inverted.
The displacement function is created in the following way:
\begin{equation}
    d(r_{\mathrm{dmo}} | M, c) = r_{\mathrm{dmb}}(M) - r_{\mathrm{dmo}}(M)
\end{equation}
for a halo with mass $M$ and concentration $c$. 

The original baryon correction model has 11 free parameters controlling $\rho_{\mathrm{clm}}$, $\rho_{\mathrm{gas}}$, and $\rho_{\mathrm{cga}}$ and $\rho_{\mathrm{2h}}$ components; see Table~1 in \citet{Schneider2019baryon}.
Out of these 14 parameters, 4 were responsible for the gas profile, 5 for stars, 3 for dark matter, and 2 for the 2-halo term. 
The original work varied 5 out of the 14 parameters. 
Out of these parameters, \citet{Giri2021emulation} found that the $M_c$ parameter, which controls the mass dependence of the gas profile, has the highest impact on lensing maps.
The model used by \citetalias{Fluri2022kids} varied only the $M_c$ parameter and additionally its redshift evolution, modeled as a power law:
\begin{equation}
    M_c = M_c^0(1+z)^\nu    
\end{equation}
with $M_c^0$ and $\nu$ as new variable parameters.
Other parameters were fixed to the best-guess model (B-avrg), which can be found in Table~2 of \cite{Schneider2019baryon}.
The fixed parameters are motivated by observed X-ray gas fractions and hydrostatic mass bias \citep{Sun2009xray,Vikhlinin2009chandra,Gonzalex2013baryon,Eckert2016xxl}.

In \cosmogrid, we release the original baryonified maps for the KiDS-1000 survey, with the Sobol grid extended with the $M_c^0$ and $\nu$ parameters, within their original prior ranges.
We also include the same model and priors for the Stage-III forecast.
However, the shell baryonification can be re-done for any set of varying parameters within that model, using the raw \nside=2048 maps and halo catalogs.

\begin{figure*}
\centering
\includegraphics[width=0.95\textwidth]{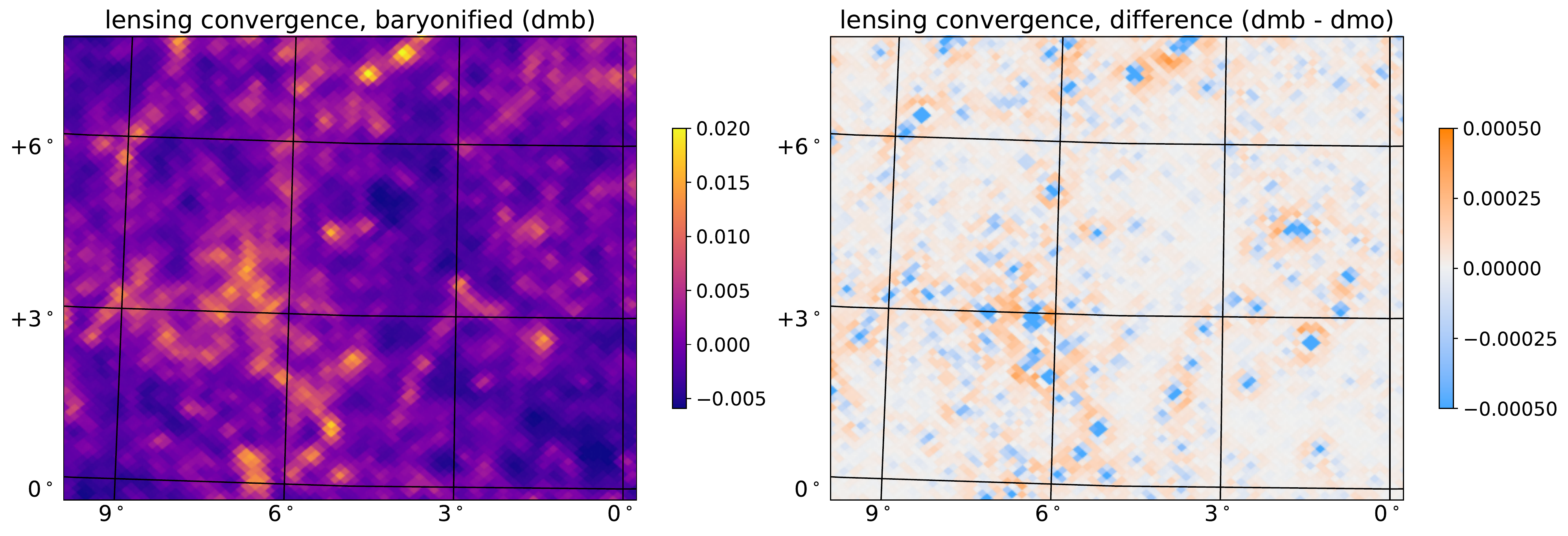}
\caption{
An example lensing convergence map with shell baryonification applied (left panel) and the difference between the baryonified and dark matter only map (right panel). 
The shell baryonification model used the fiducial parameters from \citepalias{Fluri2022kids}, with $M_c^0=13.82$ and $\nu=0$.
}
\label{fig:baryonification}
\end{figure*}

\subsection{The shell baryonification method}

The application of this method requires finding halos in the simulation, measuring their mass and concentration, calculating the displacement function, and modifying the positions of the particles belonging to that halo according to the calculated displacement.
In \citet{Schneider2020baryons1} this was done on particle snapshots, independently for every redshift. 

In \cosmogrid, we did not store snapshots and could not apply this method directly.
We therefore designed a \emph{shell baryonification} method, which approximates the original snapshot procedure using shell maps and halo catalogs.
It changes the values of high resolution particle counts maps according to the displacement function calculated from halo catalogs.
Instead of using the 3D integrated mass profile to calculate the displacement, it uses a projected, 2D mass profile.
We treat all particles in an given shell as though they were positioned at the mean redshift of the shell, thus introducing an approximation.
Then, the particle displacement is calculated on 2D gnomonic projection of the map centered on the position of the halo.

The procedure is applied as follows.
We create the halo catalogs using the friends-of-friends (FoF) halo finder available in \pkdgrav, with minimum number of particles set to 150.
The linking length used was set to 20\% of the mean particle separation.
We generate the halo catalog at each time step.
Then, we select halos with at least 100 particles within the virial radius $r\equiv r\vir$.
To obtain the mass $M\vir$ and concentration $c\vir$, we fit an NFW profile to each halo using logarithmically-spaced mass bins.
We create a halo lightcone using the same scheme as \pkdgrav\ uses to output the particle lightcone.
Halos that are close to the shell boundary, within the range of 20 Mpc/h, are included in both shells; this way it is possible for a halo to affect particles in both shells.
The 2D integrated mass profile is calculated as
\begin{equation}
    M^p_\chi(r)=2\pi \int_0^r s \int_0^{z_{\mathrm{max}}} \rho_\chi (s,z) \mathrm{d} z \mathrm{d} s
\end{equation}
with the limit $z_{\mathrm{max}}=50r$.
Then, the projected displacement function is
\begin{equation}
    d^p(r_{\mathrm{dmo}} | M, c) = r_{\mathrm{dmb}}(M^p) - r_{\mathrm{dmo}}(M^p).
\end{equation}
More details about this procedure are described in Appendix~C of \citetalias{Fluri2022kids}.
After calculating the displacement, we modify the pixels of the high resolution raw shell particle count maps (\nside=2048).
We calculate the gnomonic projection at each halo position and assing the new pixel values using linear interpolation to the displaced pixels.
This method has been validated against the 3D snapshot baryonification in \citetalias{Fluri2022kids}, we refer the reader to that work for details of these tests.
We found a very good agreement between the snapshot and shell baryonification.

The left panel in Figure~\ref{fig:baryonification} shows a lensing map for the second tomographic bin of the Stage-III redshift bin set, for the baryonified model, taken from the one of the fiducial simulations.
The right panel shows the difference between the baryonified (dmb) and dark matter only (dmo) version of the same map.
One can clearly notice that the halos become less ``peaky'' in the baryonified model, which leads to power suppression on small scales.
The mass is pushed out to the outer parts of the halo, as expected from the baryonic feedback models.

\section{Intrinsic alignments and biasing} \label{sec:intrinsic_alignments}

Intrinsic galaxy alignment is a correlation of the shape of the galaxy with structure of the density field it resides in \citep[see][for review]{Kirk2015intrinsic}.
The commonly used Non-linear Linear Alignments model \citep[NLA,][]{Hirata2004intrinsic,Bridle2007constraints,Joachimi2011ia} can be used with \cosmogrid\ by creating the IA convergence maps using the $W^{\mathrm{IA}}$ kernel shown in Equation~\ref{eqn:projection_kernels}.
This method been already used by \citepalias{Zuercher2022despeaks,Fluri2022kids} and other works. 
It has also later been described by \cite{HarnoisDeraps2022ia}.
More complicated models based on this prescription can also be designed; redshift evolution can be easily included, as well as separate parameterization for red and blue galaxies.
Including higher order terms in the Tidal Alignment Tidal Torque model \citep[TATT,][]{Blazek2019beyond} may also be possible, but was not yet demonstrated.
See \citetalias{Fluri2019kids,Zuercher2022despeaks} for more details on the map-level IA modelling and tests.
In a typical application, the IA maps can be calculated separately and combined with the lensing convergence on-the-fly.

Large scale galaxy bias describes the the statistical relation between the distribution of galaxies and matter \citep[see][for review]{Desjacques2018bias}.
The raw matter density field maps can be transformed into galaxy density fields using linear and non-linear biasing prescriptions.
The most common method is the linear galaxy bias $b$, where the galaxy number count field $\delta_g$ depends linearly on the matter density field \mbox{$\delta_g = \bar N (1+b\delta)$}, where  $\bar N$ is the average number of galaxies.
Non-linear bias models add higher order terms to this formula.
To create a forward-model for the galaxy number count maps, one can use the Poisson noise model with rate $\delta_g$ \citep[see][for example]{KacprzakFluri2022deeplss}.
Other models, including stochasticity, can also be used \citep{Gruen2018densitysplit,Friedrich2018density}.

\section{Power spectra tests} \label{sec:tests}

\begin{figure*}
\centering
\includegraphics[width=0.45\textwidth]{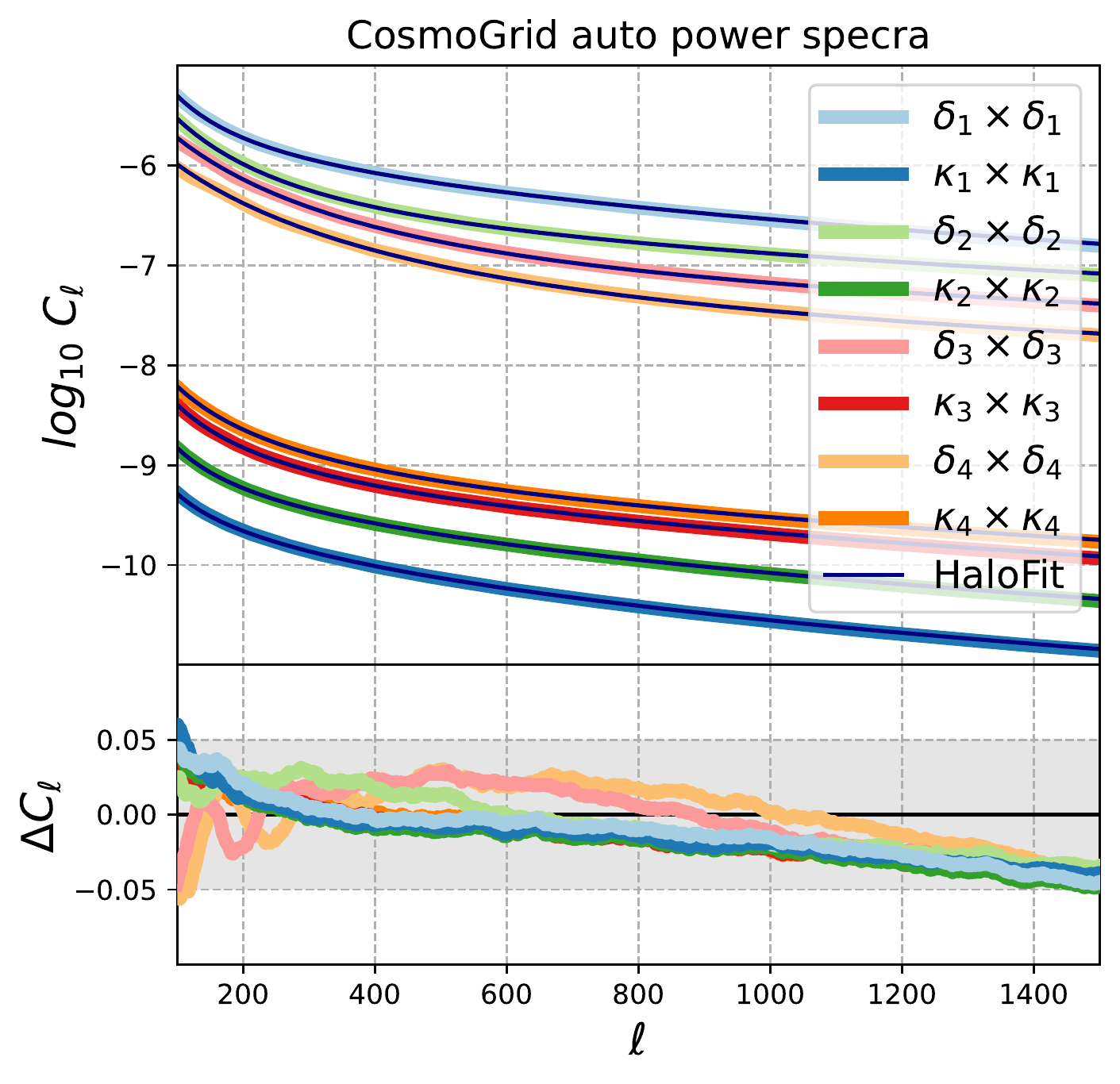}
\includegraphics[width=0.45\textwidth]{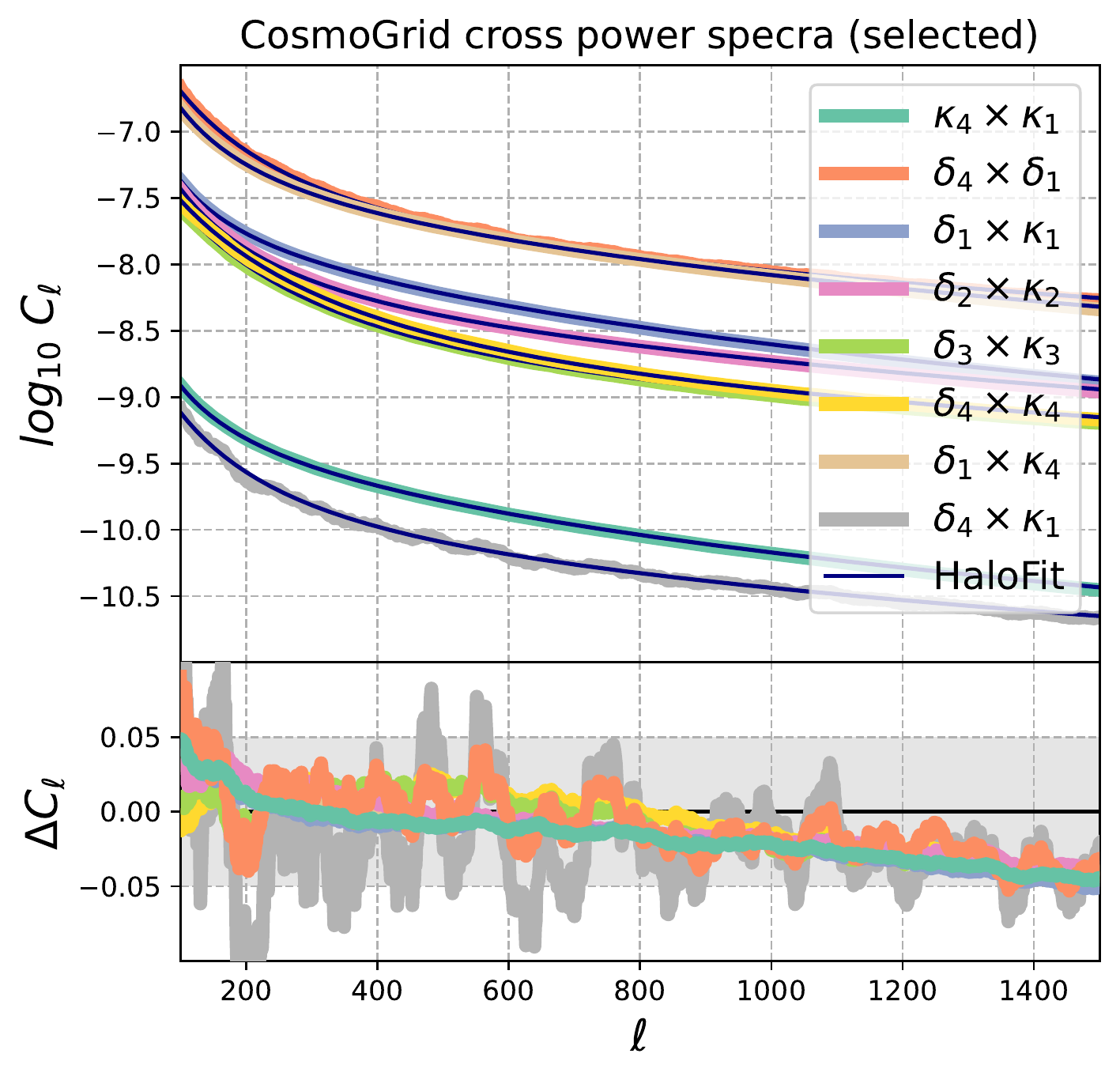}
\caption{Comparison of auto (left panel) and cross (right panel) power spectra between the noise-free lensing convergence $\kappa$ and galaxy density $\delta$ maps from the \cosmogrid\ simulations and the revised \textsc{Halofit} model \citep{Takahashi2012halofit}. 
For clarity, we show only selected eight cross spectra. 
The 5\% difference between \pkdgrav\ and \textsc{Halofit} is known and expected \citep{Tan2022nonlinear,Knabenhans2020euclidemulator2}.
The bottom panel shows the fractional difference $\Delta \cl=\cl^{\rm{CosmoGrid}}/\cl^{\rm{Halofit}}-1$.}
\label{fig:power_spectra}
\end{figure*}

We test the power spectra \cl\ calculated from the projected maps and compare them with the prediction from \pycosmo, using the revised \textsc{Halofit} method \citep{Takahashi2012halofit}.
For the KiDS-1000 maps, this test was already performed by \citetalias{Fluri2022kids}.
Here, we compare the Stage-III forecast configuration using the full sky maps that we include in the public release.
The \cl\ from simulations are computed through the spherical harmonics decomposition using the \textsc{Healpy} package.
Figure~\ref{fig:power_spectra} shows the \cl\ computed from the lensing convergence $\kappa$ and galaxy density $\delta$ maps, \rev{at \nside=2048.}.
We omit the intrinsic alignment convergence for clarity, as it uses almost the same kernel as the $\delta$ maps.
Left panels shows the auto correlations of the probes in four tomographic bins, while the right panels shows the selected cross correlations.
The lower panels show the fractional difference between the \cosmogrid\ and \textsc{Halofit} spectra $\Delta \cl=\cl^{\rm{CosmoGrid}}/\cl^{\rm{Halofit}}-1$.
\rev{The agreement is overall within 5\%, with the difference stemming from two sources: differences between \textsc{Halofit} and \pkdgrav\ \citep{Tan2022nonlinear,knabenhans2019euclid}, and the the fact that the maps are effectively smoothed by the pixel kernel, which starts to decrease the power spectrum at high $\ell$.}

\section{Testing the feature vectors against benchmarks} \label{sec:benchmarks}

\cosmogrid\ can be used for different types of non-Gaussian statistics and machine learning.
However, every analysis design should be tested to ensure that the choices made in making the simulations (i) do not introduce significant biases in the chosen features and (ii) capture the uncertainty in the features sufficiently well. 
Such tests have been demonstrated by \citetalias{Fluri2022kids,Zuercher2022despeaks}.
Here, we demonstrate how to test a chosen feature vector against \cosmogrid\ benchmarks.

We design a prototype analysis with power spectra and peak counts for tomographic weak lensing $\kappa$ and galaxy clustering $\delta$ maps.
We chose the 10 galaxies/arcmin$^2$ divided equally into tomographic bins for the Stage-III survey, as shown in Figure~\ref{fig:shells_nz}. 
The chosen survey area was 3437.75 deg$^2$, which is 1/12 of the sphere (Healpix \nside=1).
We used the same tomographic bins for lensing and clustering maps, with the same number of galaxies.
As this is a simiplified mock analysis, we neglect astrophysical effects (intrinsic alignment, baryons, magnification, redshift space distortions, and others) and measurement systematics (galaxy selection effects, shape measurement error, redshift errors, for example).

We measured the feature vectors for $\cl\in[100,1000]$ with 5 logarithmically-spaced bins, and for the peak counts for $\kappa\in[0, 0.05]$ and $\delta\in[400,700]$,  with 5 linearly spaced bins for both probes.
We assumed non-linear galaxy bias with $b=1.5$ and no intrinsic galaxy alignments.
We smoothed the probe maps with redshift bin -dependent kernels: FWHM= 4.8, 3.5, 2.8, 2.5 arcmin for $\kappa$ and FWHM=20, 16, 12, 8 arcmin for $\delta$.

\begin{figure*}
\centering
\includegraphics[width=1\textwidth]{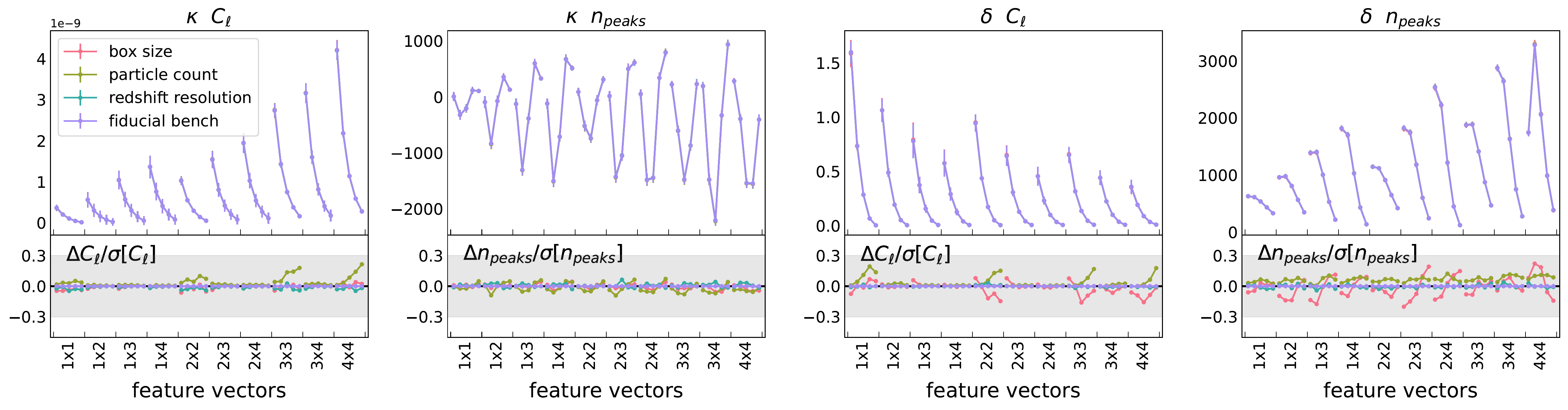}
\caption{
Mean feature vectors from different benchmarks, for the power spectra \cl\ and peak counts \npeaks, for lensing convergence $\kappa$ and galaxy clustering $\delta$.
The feature vectors are organized according to the tomographic bin combinations.
The bottom panels show the difference between the fiducial and other benchmarks, divided by the standard deviation of this feature.
Most of the differences are very small compared to their uncertainty.
}
\label{fig:features_mean}
\hspace{1cm}
\includegraphics[width=1\textwidth]{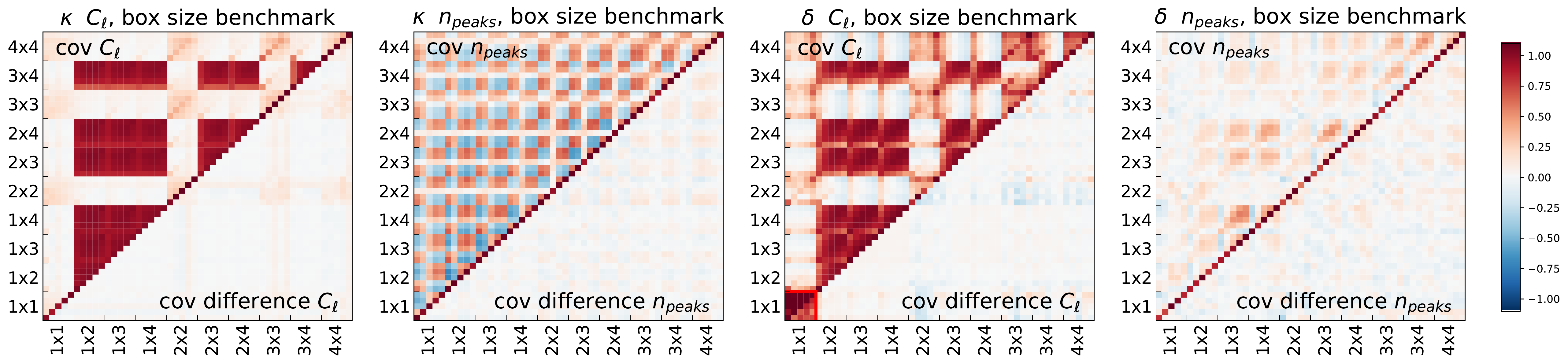}
\caption{
Pearson's correlation matrices for the chosen tomographic probes and features. 
The top triangular part shows the correlation matrix, and the diagonal and lower triangular part shows the difference between the fiducial and big box benchmark, calculated with Equation~\ref{eqn:cov_diff}.
Most of the differences are very small.
The red square for the 1\x1 bin in the clustering \cl\ marks the region where the covariance matrix does not agree well. 
}
\label{fig:features_cov}
\end{figure*}

We added Gaussian galaxy shape noise with $\sigma=0.4$ to the $\kappa$ maps and Poisson noise to the $\delta$ maps, following \citet{KacprzakFluri2022deeplss}.
We calculated the feature vector from the 200 projected full-sky maps at the fiducial reference and the benchmarks.
Those maps were created using the shell permutation scheme described in Section~\ref{sec:lightcone_permutations}, using the same random seeds for shell permutation between benchmarks.
The total number of data vectors computed was 200\x12=2400.
We compare the mean of these data vectors between the fiducial and benchmark simulations.
We also compare the covariance matrix of the feature vectors between the fiducial and the big box benchmark.
\\

Figure~\ref{fig:features_mean} shows the mean feature vectors for the power spectrum \cl\ and peak counts \npeaks, and $\kappa$ and $\delta$ probes, for the fiducial simulations and all benchmarks.
The bottom panels show the difference between the mean data vector between the fiducial and the benchmarks, divided by the standard deviation of the feature vector in each bin.
For most of the feature vector entries, these differences are subdominant to the variance, which means that the measurement with these features would not be biased.
This way the peaks and power spectra for this analysis configuration can be considered accurate enough for a measurement with \cosmogrid.

Figure~\ref{fig:features_cov} shows the Pearson's correlation matrices for the chosen features and probes for the fiducial benchmark (upper triangular), as well as its difference with respect to the big box benchmark (diagonal and lower triangular).
The difference was calculated as 
\begin{equation}
\label{eqn:cov_diff}
\frac{C^{\mathrm{fid}}_{ij}-C^{\mathrm{big}}_{ij}}{\sqrt{C^{\mathrm{fid}}_{ii} C^{\mathrm{fid}}_{jj}}}
\end{equation}
where $C^{\mathrm{fid}}$ and $C^{\mathrm{big}}$ are the covariance matrices for the fiducial and big box benchmarks, respectively.
The differences between the covariance matrices are generally smaller than 5\%, for every element, with the exception of the first clustering \cl\ for the first redshift bin. 
The covariance matrix corresponding to this part of the data vector is marked with a red square on the third panel.
This would suggest that this part of the data vector may not be suitable to be used for inference with \cosmogrid.
Further investigation could be performed to see how this difference can affect the final cosmological constraints.
Other parts of the data vector generally pass this test.

Given that this mock analysis serves just as a demonstration of the testing procedure against the benchmarks, we do not compute the forecasted cosmological constraints with these features.
The uncertainties in the mean and covariance of the data vectors can be easily propagated to the constraints.

This test has been done without cross-probe and cross-feature covariances, for simplicity.
If these are considered to be used, we would recommend to repeat this test.
Generally, these tests should be repeated for each analysis choice and target dataset.

\section{Conclusions} \label{sec:conclusions}

In this paper we introduce \cosmogrid: a large lightcone simulation set for practical map-level cosmological parameter inference from large scale structure probes.
It spans the \wcdm\ model and is aimed at Stage-III photometric datasets.
The backbone of \cosmogrid\ consists of a total of 20128 N-body runs divided into grid, fiducial and benchmark sets.
The main grid consists of 2500 unique cosmologies, each with 7 independent N-body runs. 
We introduce a novel \emph{shell permutation} scheme that avoids the box replication effects long all line of sights by combining shells from multiple simulations with unique initial conditions.

Baryon feedback can be added in post-processing using the shell baryonification method based on \citet{Schneider2019baryon}, which allows to include these effects on the map level for different models.
NLA-based intrinsic galaxy alignment models can be used to create IA convergence maps.
Linear and non-linear biasing prescriptions can be added in post-processing of galaxy density maps, similarly to \citet{KacprzakFluri2022deeplss}.

We compare \cosmogrid\ and other datasets that can be used for map-level inference in Table~\ref{tab:sims_compare}. 
Compared to the state of the art \textsc{cosmo-Slics}, we sample the full \wcdm\ space a factor of 100\x denser, while producing 7 independent simulations for each point in the parameter space.
\cosmogrid, however, has lower particle count and therefore is not as precise as \textsc{cosmo-Slics} at resolving small scales. 

We show the full parameters used for different \cosmogrid\ simulation sets in Table~\ref{tab:sim_params} and explain the design choices in creating the simulations.
We explain the trade-offs between bias and variance with respect to computational resources in Section~\ref{sec:sims}.
While \cosmogrid\ provides predictions that are accurate enough for the Stage-III statistics investigated, a careful study of the impact of these choices should be performed for every new proposed analysis.
Depending on the sets of features used, the approximations used in \cosmogrid\ can potentially introduce some parameter bias, or result in too large errors on feature covariance.
To test this in a robust way, we also include a set of specialized benchmark simulations.
Such tests were already performed by \citetalias{Fluri2022kids,Zuercher2022despeaks}.
In this work, we present a mock analysis with tomographic power spectra \cl\ and peak counts \npeaks, for lensing convergence $\kappa$ and galaxy clustering $\delta$ maps.
We show how to use the benchmarks to assess the bias and covariance errors induced by the approximations in \cosmogrid\ for a target Stage-III foracast experiment.
These tests should be repeated for each new analysis type and target dataset.

The full content of the \cosmogrid\ data release is described in Section~\ref{sec:cosmogrid_intro}.
The raw particle count maps can be used to create probe maps for a new survey. 
This can be followed by baryonification, IA and biasing modelling.
As this process can be quite complex and computationally demanding, we create projected maps for a Stage-III forecast configuration with 4 redshift bins, shown in the left panel of Figure~\ref{fig:shells_nz}.
This dataset is much smaller and should be more convenient to use for quick forecasting of new analysis types.
We also include the original lensing and intrinsic alignments maps used by \citetalias{Fluri2022kids} in the KiDS-1000 analysis.

\rev{While \cosmogrid\ was designed for Stage-III surveys, one may consider its use for Stage-IV photometric surveys, such as LSST or Euclid.
Those should be approached with caution, for several reasons.
Firstly, the precision of parameter constraints from Stage-IV will be several times better; the uncertainties from SBI with \cosmogrid\ in that regime may not be accurate enough, due to the box replication strategy.
Secondly, the baryon feedback models are unlikely to be sufficient at that level of precision.
Finally, at the time of Stage-IV, it may be more practical to use a narrower prior grid, which would enable denser sampling in the 5$\sigma$ region of the expected contours.
}

\begin{acknowledgments}
This work was supported by a grant from the Swiss National Supercomputing Centre (CSCS) under project ID s998.
The CSCS large production project was called ``Measuring Dark Energy with Deep Learning'' (PI: Tomasz Kacprzak).
JF would like to thank Jeppe Mosgaard Dakin for helpful discussions and especially his contribution to the \wzero - \omatter\ prior of the simulation grid. 
We thank Douglas Potter for helpful discussions.
AS acknowledges support from the Swiss National Science Foundation via the grant PCEFP2 181157.
We thank Alexander Reeves for help with the \textsc{UFalcon} package.
We thank Aurelien Lucchi for ongoing collaboration on machine learning application in cosmology.
We would like to thank the technical support teams of the Euler and Piz Daint computing clusters.
We thank Christian Herzog from ETH Zurich Physics ISG for building the hard drive hosting \cosmogrid, and Christian Bolliger for help with setting up the Globus endpoint.
TK thanks Peter Melchior for creating the publicly available \textsc{SkyMapper} python package for plotting spherical maps.
\end{acknowledgments}

\bibliography{bibliography.bib}{}
\bibliographystyle{aasjournal}

\end{document}